\documentclass[final,5p,times,twocolumn]{elsarticle}

\usepackage[utf8]{inputenc}

\usepackage{amssymb}
\usepackage{amsmath}
\usepackage{comment}
\usepackage{enumitem}
\usepackage{listings}
\newcommand{\calR}{{\cal R}}
\newcommand{\calS}{{\cal S}}

\newcounter{bla}

\journal{Computer Physics Communications}

\begin{document}

\begin{frontmatter}



\title{CoOMBE: A suite of open-source programs for the integration of the\\ optical Bloch equations and Maxwell-Bloch equations}


\author[a]{R. M. Potvliege\corref{cor1}}
\author[a]{S. A. Wrathmall\corref{cor2}}

\cortext[cor1]{Corresponding author. \textit{E-mail address:} r.m.potvliege@durham.ac.uk}
\cortext[cor2]{\textit{E-mail address:} s.a.wrathmall@durham.ac.uk}
\address[a]{Department of Physics, Durham University, South Road, Durham DH1~3LE, UK}

\begin{abstract} 
The programs described in this article and distributed with it aim (1) at integrating the optical Bloch equations governing the time evolution of the density matrix representing the quantum state of an atomic system driven by laser or microwave fields, and (2) at integrating the 1D Maxwell-Bloch equations for one or two laser fields co-propagating in an atomic vapour. The rotating wave approximation is assumed. 
These programs can also be used for more general quantum dynamical systems governed by the Lindblad master equation. They are written in Fortran~90; however, their use does not require any knowledge of Fortran programming. Methods for solving the optical Bloch equations in the rate equations limit, for calculating the steady-state density matrix and for formulating the optical Bloch equations in the weak probe approximation are also described.





\end{abstract}


\end{frontmatter}



\noindent {\bf PROGRAM SUMMARY}

\begin{small}
\noindent
{\em Program Title:} CoOMBE\\
{\em CPC Library link to program files:} (to be added by Technical Editor) \\
{\em Developers' repository link:} https://github.com/durham-qlm/CoOMBE \\
{\em Code Ocean capsule:} (to be added by Technical Editor)\\
{\em Licensing provisions:} GPLv3 \\
{\em Programming language:} Fortran~90                                  \\
{\em Nature of problem:}
The present programs can be used for the following operations: (1) Integrating the optical-Bloch equations within the rotating wave approximation for a multi-state atomic system. At the choice of the user, the calculation will return either the time-dependent density matrix at given times or the density matrix in the long time limit if the system evolves into a steady state in that limit. The calculation can be done with or without averaging over the thermal velocity distribution of the atoms. The number of atomic states which can be included in the calculation is limited only by the CPU time available and possibly by memory requirements. An arbitrarily large number of laser or microwave fields can be included in the calculation if these fields are all CW. This number is currently limited to one or two for fields that are not all CW.
The calculation can be done in the weak probe approximation, or in the rate equations approximation, or without assuming either of these two approximations. Calculating refractive indexes, absorption coefficients and complex susceptibilities is also possible. (2) Integrating the 1D Maxwell-Bloch equations in the slowly varying envelope approximation for one or two fields co-propagating in a single-species atomic vapour.
Although geared towards the case of atoms interacting with laser fields, this code can also be used for more general quantum systems with similar equations of motion (e.g., molecular systems, spin systems, etc.).\\ 
{\em Solution method:}
The Lindblad master equation is expressed as a system of homogeneous first order linear differential equations, which are transformed as required and solved to obtain the density matrix representing the state of the atomic system.
A variety of methods are offered to this end. The same approach is also used in the calculation of the polarisation of the medium when integrating the Maxwell-Bloch equations. The latter are integrated over space using predictor-corrector methods. The library includes a general driving program making it possible to use these codes without additional program development. The distribution also includes examples of the use of a container for running these programs without a pre-installed Fortran compiler.
\\
   \\

\end{small}


\section{Introduction}
\label{section:intro}

The programs described in this article have been developed for modelling general atomic or molecular systems interacting with one or several laser or microwave fields resonant or nearly resonant with atomic transitions, the interaction being treated within the rotating wave approximation. They can also be used for more general quantum systems, e.g., spin systems, governed by similar equations of motion. Their main focus is on the calculation of the density matrix representing the state of the system as obtained by integrating the optical Bloch equations (i.e., the Lindblad master equation for such systems). The populations and coherences can be calculated as time-dependent functions. Alternatively, for systems driven by CW fields and evolving to a steady state, they can also be obtained in the long time limit. Calculations of complex susceptibilities, refractive indexes and absorption coefficients are also possible. 
The present programs were originally developed for studying the formation of optical solitons and two-colour quasisimultons in an optical vapour, which required the integration of the 1D Maxwell-Bloch equations in the slowly envelope approximation \cite{Ogden2019}. The Maxwell-Bloch solvers developed at that occasion are included in this library as they are of general interest and share the same user interface.\footnote{Python codes developed in the course of the work reported in \cite{Ogden2019} are published at the URL https://github.com/tpogden/maxwellbloch.}

A number of open source programs are already available for tackling similar or related calculations, namely general purpose programs for the modelling of open quantum systems, programs more specific to Atomic Physics calculations, and programs solving the Maxwell-Bloch equations in various approximations. The general purpose programs include, in particular, Qutip \cite{Johansson2012,Johansson2013}, written in Python, its predecessor, Quantum Optics Toolbox \cite{Tan1999,Tan1999b}, written in MATLAB, and a more recent alternative, QuantumOptics, written in Julia \cite{Kraemer2018}. They also include several MATLAB programs primarily intended for educational purposes \cite{Korsch2016,Norembuena2020,Meher2024}, two Quantum Monte Carlo programs written in C++ \cite{Schack1997,Vukics2007} and a MATLAB program focusing on the optimal control of the dynamics of quantum systems interacting with external electromagnetic fields \cite{Schmidt2018}. General programs solving the optical Bloch equations for atomic systems have also been published, including the Atomic Density Matrix package \cite{ADM}, which is written in Mathematica and also supports more general quantum optical calculations, a collection of Python tools for modelling few-level atom-light interactions \cite{Downes2023}, and PyLCP \cite{Eckel2022}, a Python program oriented towards the modelling of laser cooling but also allowing for general solutions of the optical Bloch equations. The Elecsus program, also written in Python, is specialised to the case of an atomic vapour addressed by a single probe field and offers powerful facilities for the analysis of experimental absorption spectra for that particular case \cite{Zentile2015,Keaveney2018}. The necessary atomic data are provided by Atomic Density Matrix, PyLCP and Elecsus for species of current interest. The Maxwell-Bloch solvers include mbsolve \cite{Riesch2021}, a C++ program for the 1D propagation of a field in the plane wave approximation or a field confined to a wave guide, and QuEST \cite{Glosser2021,Lu2022,Lu2023,QuEST}, also written in C++, which was developed for modelling the interaction of an electromagnetic field with multiple 2-level quantum dots in 3D. Neither mbsolve nor QuEST assume the slowly varying envelope approximation; mbsolve does not assume the rotating wave approximation either and is not restricted to a 2-state medium (the Maxwell-Bloch solver included in the present package is simpler: it takes the medium to be homogeneous and assumes 
both the slowly varying envelope approximation and the rotating wave approximation, which is appropriate, e.g., in calculations of self-induced transparency for many-cycle pulses).

The present programs may nonetheless be of interest in view of their generality, their scalability to large systems, and the wide choice of integration methods they offer. While written in Fortran~90 for speed and convenience, no knowledge of this language is necessary for using them. It is expected that they will be further extended in the future, in particular by coupling them with programs providing the atomic data required for calculations on atomic systems of current experimental interest. Future versions will be published at the URL https://github.com/durham-qlm/CoOMBE, a GitHub repository of the Quantum, Light and Matter research group of Durham University.

General information about the distribution is given in Section~\ref{section:general}. The computational methods implemented in these program and the theoretical framework are outlined in Section~\ref{section:theory}. Information about installing and using this software and information can be found in Section~\ref{section:using_honours} and also, at much greater lengths, in the User Manual. The reuse of codes written by other authors in the present programs is acknowledged in Section~\ref{section:notice}. The main text is accompanied by a number of more technical appendices: the reduction of Lindblad master equation to rate equations is explained \ref{appendix:rate}, the calculation of a steady-state density matrix in the long time limit in \ref{appendix:steady}, and the implementation of the weak probe approximation in the present computational framework in \ref{appendix:weak}. Useful results concerning weak probe calculations for a single field are gathered in \ref{appendix:single}. Examples of the use of these programs are given in \ref{appendix:example} and \ref{appendix:example2}, and further examples can be found in the {\tt examples} directory included in the distribution. Advice about how to run this software without the installation of a Fortran compiler and supporting libraries,  through a Podman container \cite{podman}, are provided in \ref{appendix:podman} and in the GitHub repository.

We are not aware that the methods described in \ref{appendix:rate}, \ref{appendix:steady} and \ref{appendix:weak} are widely known or previously published.

\section{General information}
\label{section:general}

\subsection{Organisation into program units}

This library contains several modules and external subroutines, as follows:
\begin{itemize}
\item The {\tt general\_settings} modules. This module sets several key parameters, in particular the variable ${\tt nst}$  which defines the number of states to be considered in the calculation. More information about this module and these parameters can be found in Section~\ref{section:keyparameters}.
\item The {\tt obe\_constants} module, which defines fundamental physical constants used elsewhere in the code \cite{Tiesinga2018}. 
\item The {\tt obe} module, which forms the main part of the library. It contains a number of subprograms, many of which are private to this module (i.e., cannot be called from outside the module). These subprograms are concerned with solving the optical Bloch equations and/or forming a user-friendly interface with the {\tt ldbl} module. 
\item The {\tt mbe} module, grouping program units concerned with solving the Maxwell-Bloch equations and with solving the optical Bloch equations for time-dependent fields.
\item The {\tt ldbl} module, which contains a number of subprograms concerned with setting up and solving the Lindblad master equation. This module is the core of the library. However, the subprograms it contains can be accessed more conveniently through subprograms forming part of the {\tt obe} or {\tt mbe} modules, and using those does not require any knowledge of the inner working of the ${\tt ldbl}$ module. For this reason, the latter is not addressed in the present article. The reader is referred to the detailed documentation for general information about its content.
\item The external subroutine {\tt ext\_setsys}, which is used only for communicating information between the {\tt obe} and {\tt mbe} modules.
\item The external subroutines {\tt fcn\_dummy} and {\tt solout\_dummy}, which are provided for compatibility with the original code of the DOP853 ODE solver mentioned in Section~\ref{section:timedep}.
\item The {\tt ldblstore} module, which is used to store certain intermediate results produced by programs contained in the {\tt ldbl} module.
\item The {\tt driveall} program, described in Section~\ref{section:driveall}, which offers a simple interface with {\tt obe} and {\tt mbe} and makes it possible to use these codes without any Fortran programming.
\end{itemize}
These various components are grouped into files as outlined in Table~\ref{table:files}.

Besides a number of program units intended for internal use only, the {\tt obe} and {\tt mbe} modules currently contain a total of 35 user-facing subprograms (Table~\ref{table:obembe_contents}), i.e., subprograms providing an interface between the internal program units and a user-written driving program.

\subsection{Documentation and examples}

The distribution includes a User Manual complementing the present article. This document contains further information about all the programs units forming this library, a detailed description of the user-facing subprograms, detailed information about the use of the {\tt driveall} program, and a short tutorial explaining how the Hamiltonian of an atomic system interacting with an electromagnetic field treated in the rotating wave approximation can be cast into the form of Eq.~(\ref{eq:Hprime}).

Further examples illustrating the use of various features of this library are also provided. The corresponding files and documentation can be found in the {\tt examples} directory.

\begin{table}[h!]
\caption{\label{table:files}Contents of the distribution}
\centering
\footnotesize
\begin{tabular}{l l}
{}\\[-2mm]
\hline\\ { }\\[-6mm] 
{File or directory} & Short description\\
{ }\\[-3mm]
\hline\\
{}\\[-5mm]
{\tt user\_manual.pdf}
& 
\parbox[t]{5cm}{The User Manual, including a detailed description of the use of the {\tt driveall} program.}\\[-3mm]
\\
{\tt general\_settings.f90}
&\parbox[t]{5cm}{The {\tt general\_settings} module.}\\[-3mm]
\\
{\tt ldbl.f90}
&\parbox[t]{5cm}{The {\tt ldbl} and {\tt ldblstore} modules and the {\tt fcn\_dummy} and {\tt solout\_dummy} subroutines.}\\[-3mm]
\\
{\tt obe.f90}
&\parbox[t]{5cm}{The {\tt obe} and {\tt obe\_constants} modules and the {\tt ext\_setsys} subroutine.}\\[-3mm]
\\
{\tt mbe.f90}
&\parbox[t]{5cm}{The {\tt mbe} module.}\\[-3mm]
\\
{\tt driveall.f90}
&\parbox[t]{5cm}{The {\tt driveall} program.}\\[-3mm]
\\
{\tt examples}
&
\parbox[t]{5cm}{Examples of the use of this software, including the files and the program listed in \ref{appendix:example} and \ref{appendix:example2}.}\\[-3mm]
\\
{ }\\[-2mm]
\hline 
\end{tabular}
\end{table}

\begin{table*}[h!]
\caption{\label{table:obembe_contents}User-facing subprograms contained in the \texttt{obe} and \texttt{mbe} modules.}
\centering
\footnotesize
\begin{tabular}{l l}
{}\\[-2mm]
\hline\\ { }\\[-6mm] 
\parbox{4.6cm}{Name} & Short description\\
{ }\\[-3mm]
\hline\\
{}\\[-4mm]
\multicolumn{2}{l}{\underline{Computational routines: Time-dependent calculation of the density matrix}} \\
{}\\[-2mm]
{\tt mbe\_tdint\_1}
&\parbox[t]{12.7cm}{
For a single field with a time-dependent envelope, with or without Doppler averaging.}\\[-3mm]
\\
{\tt mbe\_tdint\_2}
&\parbox[t]{12.7cm}{
For a superposition of two fields with a time-dependent envelope, with or without Doppler averaging.}\\[-3mm]
\\
{\tt obe\_Doppler\_av\_td\_A}
&\parbox[t]{12.7cm}{
For CW fields, with Doppler averaging.}\\[-3mm]
\\
{\tt obe\_Doppler\_av\_td\_B}
&\parbox[t]{12.7cm}{
For CW fields, with Doppler averaging 
 (alternative method).}\\[-3mm]
\\
{\tt obe\_tdint}
&\parbox[t]{12.7cm}{
For CW fields, without Doppler averaging.}\\[-3mm]
\\[3mm]
\multicolumn{2}{l}{\underline{Computational routines: Calculation of the steady state density matrix for CW fields}} \\
{}\\[-2mm]
{\tt obe\_2state}
&\parbox[t]{12.7cm}{
For 2-state systems driven by a single field, with or without Doppler averaging.}\\[-3mm]
\\
{\tt obe\_Doppler\_av\_st}
&\parbox[t]{12.7cm}{
For general systems with semi-analytical Doppler averaging.}\\[-3mm]
\\
{\tt obe\_Doppler\_av\_st\_numerical}
&\parbox[t]{12.7cm}{
For general systems with numerical Doppler averaging.}\\[-3mm]
\\
{\tt obe\_steadystate}
&\parbox[t]{12.7cm}{
For general systems without Doppler averaging.}\\[-3mm]
\\
{\tt obe\_steadystate\_ladder}
&\parbox[t]{12.7cm}{
For ladder systems in the weak probe approximation, with or without Doppler averaging.}\\[-3mm]
\\
{\tt obe\_steadystate\_onefld}
&\parbox[t]{12.7cm}{
For one-field systems, with or without Doppler averaging (intended for calculation repeated for multiple values of the detuning).}\\[-3mm]
\\
{\tt obe\_steadystate\_onefld\_powerbr}
&\parbox[t]{12.7cm}{
As {\tt obe\_steadystate\_onefld\_weakprb} but with power broadening taken into account.}\\[-3mm]
\\
{\tt obe\_steadystate\_onefld\_weakprb}
&\parbox[t]{12.7cm}{
For one-field systems in the weak probe approximation, with or without Doppler averaging.}\\[-3mm]
\\
{\tt obe\_weakfield}
&\parbox[t]{12.7cm}{
For one-field systems in the weak probe approximation, with or without Doppler averaging (with calculation of the refractive index and the absorption coefficient).}\\[-3mm]
\\
{\tt obe\_weakprb\_3stladder}
&\parbox[t]{12.7cm}{
For 3-state ladder systems in the weak probe approximation, with or without Doppler averaging.}\\[-3mm]
\\
{\tt obe\_weakprb\_4stladder}
&\parbox[t]{12.7cm}{
For 4-state ladder systems in the weak probe approximation, with or without Doppler averaging.}\\[-3mm]
\\[3mm]
\multicolumn{2}{l}{\underline{Computational routines: Integration of the Maxwell-Bloch equations}} \\
{}\\[-2mm]
{\tt mbe\_propagate\_1}
&\parbox[t]{12.7cm}{
Propagates a single field.}\\[-3mm]
\\
{\tt mbe\_propagate\_2}
&\parbox[t]{12.7cm}{
Propagates a superposition of two fields.}\\[-3mm]
\\[3mm]
\multicolumn{2}{l}{\underline{Initialisation routines}} \\
{}\\[-2mm]
{\tt mbe\_set\_envlp}
&\parbox[t]{12.7cm}{
Defines the parameters of the temporal envelope of either the probe or the coupling field.}\\[-3mm]
\\
{\tt mbe\_set\_tdfields\_A}
&\parbox[t]{12.7cm}{
Calculates the temporal envelope of either the probe or the coupling field.}\\[-3mm]
\\
{\tt mbe\_set\_tdfields\_B}
&\parbox[t]{12.7cm}{
Passes a time mesh and pre-calculated time-dependent complex amplitude(s) of the applied field(s) to {\tt mbe}.}\\[-3mm]
\\
{\tt obe\_reset\_campl}
&\parbox[t]{12.7cm}{
Sets the complex amplitude of a field to a new value.}\\[-3mm]
\\
{\tt obe\_reset\_detuning}
&\parbox[t]{12.7cm}{
Sets the detuning of a field to a new value.}\\[-3mm]
\\
{\tt obe\_setcsts}
&\parbox[t]{12.7cm}{
Sets a number of parameters defining the system considered.}\\[-3mm]
\\
{\tt obe\_set\_Doppler}
&\parbox[t]{12.7cm}{
Sets parameters related to Doppler averaging.}\\[-3mm]
\\
{\tt obe\_setfields}
&\parbox[t]{12.7cm}{
Specifies field parameters.}\\[-3mm]
\\
{\tt obe\_setoutputfiles}
&\parbox[t]{12.7cm}{
Specifies the unit numbers of output files.}\\[-3mm]
\\
{\tt obe\_set\_tol\_dop853}
&\parbox[t]{12.7cm}{
Specifies convergence criteria for the DOP853 ODE solver.}\\[-3mm]
\\[3mm]
\multicolumn{2}{l}{\underline{Auxiliary routines}} \\
{}\\[-2mm]
{\tt obe\_coher\_index}
&
\parbox[t]{12.7cm}{
Given the indexes of two different atomic states, $i$ and $j$, returns the indexes of the components corresponding to the real and imaginary parts of the coherence $\rho_{ij}$ in the 1D representation of the density matrix.}\\[-3mm]
\\
{\tt obe\_fieldtocfield}
&\parbox[t]{12.7cm}{
Given a variable of type {\tt obefield} containing the details of a field, returns a variable of type {\tt obecfield} containing the same details.}\\[-3mm]
\\
{\tt obe\_find\_campl}
&\parbox[t]{12.7cm}{
Given a complex dipole moment and a complex Rabi frequency, returns the corresponding complex electric field amplitude.}\\[-3mm]
\\
{\tt obe\_find\_rabif}
&\parbox[t]{12.7cm}{
Given a complex dipole moment and a complex electric field amplitude, returns the corresponding complex Rabi frequency.}\\[-3mm]
\\
{\tt obe\_init\_rho}
&\parbox[t]{12.7cm}{
Returns the density matrix of a mixed state with given populations and zero coherences.}\\[-3mm]
\\
{\tt obe\_pop\_index}
&\parbox[t]{12.7cm}{
Returns the index of the component corresponding to a specified population in the 1D representation of the density matrix.}\\[-2mm]
\\
{\tt obe\_susceptibility}
&\parbox[t]{12.7cm}{
Given the relevant coherences, calculates the complex susceptibility, refractive index and absorption coefficient.}\\[-3mm]
\\
{ }\\[-2mm]
\hline 
\end{tabular}
\end{table*}

\section{Theory and methods}
\label{section:theory}

\subsection{The optical-Bloch equations} 

\subsubsection{General formulation}
\label{section:generalformulation}

The {\tt obe} and {\tt mbe} codes have been developed for modelling atomic systems driven by laser fields or other coherent electromagnetic fields and composed of two or more atomic states, the fields being resonant or close to resonance with transitions between these states. The codes can also be used to calculate the density matrix for more general
$N$-state quantum systems interacting with a superposition of $M$ electromagnetic fields, as long as the rotating wave approximation can be assumed.

Each field is described by a real electric field vector, ${\bf E}_\alpha({\bf r},t)$, the total electric field of the applied light at position ${\bf r}$ and time $t$ being ${\bf E}({\bf r},t)$, with
\begin{equation}
    {\bf E}({\bf r},t) = 
\sum_{\alpha = 1}^M \,{\bf E}_\alpha({\bf r},t).
\label{eq:Etot}
\end{equation}
The calculation assumes that each of the ${\bf E}_\alpha({\bf r},t)$'s can be written as the product of a slowly-varying envelope and a plane-wave carrier. 
Specifically,
\begin{align}
{\bf E}_\alpha({\bf r},t) &= \frac{1}{2}\, \hat{\text{\boldmath{$\epsilon$}}}_\alpha \, {\cal E}_\alpha
\exp[i({\bf k}_\alpha\cdot{\bf r} - \omega_\alpha t)] + \mbox{c.c.} 
\label{eq:Efirst}\\
&= |{\cal E}_\alpha|\,\mbox{Re}\left(
\hat{\text{\boldmath{$\epsilon$}}}_\alpha
\exp[i({\bf k}_\alpha\cdot{\bf r} - \omega_\alpha t + \arg {\cal E}_\alpha)]
\right),
\label{eq:Esecond}
\end{align}
where ${\bf k}_\alpha$ is the wave vector of field $\alpha$ and $\omega_\alpha$ is its angular frequency ($\omega_\alpha > 0$). The field amplitudes ${\cal E}_\alpha$ may be complex and may vary in time. The polarisation vectors are assumed to be constant and of unit norm:
\begin{equation}
    \hat{\text{\boldmath{$\epsilon$}}}^*_\alpha \cdot \hat{\text{\boldmath{$\epsilon$}}}_\alpha = 1.
\end{equation}
Given Eqs.~(\ref{eq:Efirst}) and (\ref{eq:Esecond}), the intensity of a continuous wave (CW) field is related to its complex amplitude by the following equation:
\begin{equation}
I = \epsilon_0 c\,|{\cal E}|^2/2.
\end{equation}
This relation generalizes to the case of a pulsed field of envelope ${\cal E}(t)$, provided the pulse encompasses more than a few optical cycles:  $\epsilon_0 c \,|{\cal E}(t)|^2/2$ can be taken to be the instantaneous intensity at time $t$. Intensities are easily converted into electric field amplitudes and conversely by making use of the fact that an intensity of exactly 1~mW~cm$^{-2}$ corresponds to an electric field amplitude of $86.8021$~V~m$^{-1}$.

The states coupled to each other by the field(s) are assumed to be orthonormal eigenstates of the field-free Hamiltonian, $\hat{H}_0$. We denote these states by the ket vectors $|i\rangle$, $i=1,\ldots,N$, and the corresponding field-free eigenenergies by $\hbar \omega^{(i)}$: $\langle i | j \rangle = \delta_{ij}$ and
\begin{equation}
\hat{H}_0 |i\rangle= \hbar \omega^{(i)} |i\rangle, \qquad i = 1,\ldots,N.
\end{equation}
Typically,
these $N$ states form two or more groups  differing considerably in energy
and each of the fields is resonant or close to resonance with transitions between states of one of these groups and states of one of the other groups.
E.g., in rubidium, these groups could be the $5{\rm S}_{1/2}(F,m_F)$ states, the $5{\rm P}_{1/2}(F,m_F)$ states, the $5{\rm P}_{3/2}(F,m_F)$ states, etc., and the calculation could involve a field resonant or nearly resonant on a transition between one of the $5{\rm S}_{1/2}(F,m_F)$ states and one of the $5{\rm P}_{1/2}(F,m_F)$ states. 
These groups of energetically close states are denoted by ${\cal G}_1$, ${\cal G}_2$, \ldots, ${\cal G}_K$ in the following. The states belonging to a same group may or may not differ in energy, depending on the system. Either way, the energies $\hbar \omega^{(i)}$ of the states belonging to a same group can be referred to a reference energy, $\hbar\omega_{\rm ref}$, from which each one differs by an energy offset $\hbar \delta \omega^{(i)}$: for group~$k$,
\begin{equation}
    \omega^{(i)} = \omega_{\rm ref}(k) + \delta \omega^{(i)} \qquad \mbox{if}\; i \in {\cal G}_k.
    \label{eq:offsets}
\end{equation}
A reference energy $\hbar \omega_{\rm ref}(k)$ could be, for example, the energy of one of the basis states, or the centroid of a group of hyperfine levels. More generally, $\hbar \omega_{\rm ref}(k)$ can be any energy reference appropriate for the problem at hand. 

The calculation also assumes that the interaction with the fields is taken into account within the electric dipole approximation, which amounts to neglecting the spatial variation of ${\bf E}({\bf r},t)$. For simplicity, the vector ${\bf r}$ is taken to be zero in the following. The $\exp(\pm i\,{\bf k}_\alpha\cdot{\bf r})$ phase factors can be subsumed into the complex amplitudes ${\cal E}_\alpha$ should they be relevant. The Hamiltonian of the system thus takes on the following form:
\begin{align}
    \hat{H}(t) = \sum_{i=1}^N \hbar \omega^{(i)} |i\rangle\langle i| -& \frac{1}{2}\sum_{\alpha = 1}^M\sum_{i=1}^N\sum_{j=1}^N \left[
    {\cal E}_\alpha \exp(-i\omega_\alpha t)
    \langle i | \hat{\text{\boldmath{$\epsilon$}}}_\alpha \cdot \hat{\bf D} | j \rangle \right. \nonumber \\
    &\quad\left. \;+ {\cal E}_\alpha^* \exp(i\omega_\alpha t)
    \langle i | \hat{\text{\boldmath{$\epsilon$}}}^*_\alpha \cdot \hat{\bf D} | j \rangle \right] |i\rangle\langle j|,
    \label{eq:fullH}
\end{align}
where $\hat{\bf D}$ is the atom's dipole operator. In terms of the relevant position operator, $\hat{\bf X}$,
\begin{equation}
    \hat{\bf D} = -e \hat{\bf X},
\end{equation}
where $e$ is the absolute charge of the electron ($e > 0$). The matrix elements of the operator
$\hat{\text{\boldmath{$\epsilon$}}}_\alpha \cdot \hat{\bf X}$ would typically be obtained as the product of a reduced matrix element and an angular factor.
The corresponding complex Rabi frequencies $\Omega_{\alpha;ij}$ are defined as follows throughout the code: 
\begin{equation}
    \Omega_{\alpha;ij} = 
    \begin{cases}
{\cal E}_\alpha\,\langle\, i\, |\,
\hat{\text{\boldmath{$\epsilon$}}}_{\alpha} \cdot \hat{\bf D} \,|\, j\, \rangle/\hbar & \mbox{if}\;\; \hbar \omega^{(i)} > \hbar \omega^{(j)},\\
{\cal E}_\alpha^*\,\langle \, i\,|\,
\hat{\text{\boldmath{$\epsilon$}}}_{\alpha}^* \cdot \hat{\bf D}\, |\, j\, \rangle/\hbar & \mbox{if}\;\;\hbar \omega^{(i)} < \hbar \omega^{(j)},
\end{cases}
\label{eq:Omegadef}
\end{equation}
with the convention that $\Omega_{\alpha;ij} = 0$ if states $i$ and $j$ are deemed not to be coupled by field~$\alpha$, e.g., because this transition would be excessively far from resonance.
It should be noted that this definition of the complex Rabi frequency includes the (negative) $-e$ factor multiplying the position operator. It may differ, in sign and otherwise, from the definition of the Rabi frequency used by other authors.

This Hamiltonian cannot be treated in its full complexity by the present software. Rather, the {\tt obe} and {\tt mbe} routines are based on a simplified Hamiltonian, $\hat{H}'$, derived from $\hat{H}(t)$ by neglecting any excessively far detuned transition, making the rotating wave approximation and passing to slowly varying variables by a unitary transformation. 
This transformed Hamiltonian is assumed to have the following general form:
\begin{equation}
\hat{H}' = \hbar \sum_{i=1}^N 
\left(\delta \omega^{(i)} + \sum_{\alpha=1}^M a_{i\alpha} \Delta_\alpha \right) |i\rangle\langle i| - (\hbar/2) \sum_{i=1}^N\sum_{j=1}^N\Omega_{ij}
|i\rangle\langle j|,
\label{eq:Hprime}
\end{equation}
where $\Delta_\alpha$ is the frequency detuning of field $\alpha$,  the $a_{i\alpha}$'s are numerical factors, and
\begin{equation}
    \Omega_{ij} = \sum_{\alpha = 1}^M \Omega_{\alpha;ij}.
    \label{eq:Omegatot}
\end{equation}
What the factors $a_{i\alpha}$ are and how the frequency detunings are defined in terms of the energies of the relevant states and the angular frequencies $\omega_\alpha$ varies from system to system, as is explained in Appendix~A of the User Manual. For instance, for the 3-state system considered in \ref{appendix:example}, 
\begin{align}
    \Delta_1 &= \omega_1 - (\omega_{\rm ref}{(2)} - \omega_{\rm ref}{(1)}),
    \label{eq:delta1}\\
    \Delta_2 &= \omega_2 - (\omega_{\rm ref}{(3)} - \omega_{\rm ref}{(2)}),
    \label{eq:delta2}
\end{align}
and, as can be seen from Eq.~(\ref{eq:Hprimeladder}), $a_{11}=a_{12}=a_{22}=0$ and $a_{21}=a_{31}=a_{32}=-1$. We stress that these frequency detunings, as defined, are angular frequencies, like the Rabi frequencies $\Omega_{ij}$.
It can be noted that $\hat{H}'$ is a self-adjoint operator, as expected, since $\Omega_{ji} = \Omega_{ij}^*$ within the above definition of the Rabi frequencies.
For most systems, $\hat{H}'$ is constant in time if all the fields considered are CW fields.

The optical Bloch equations are the equations of motion for the individual components of the density matrix for an open quantum system interacting with classical electromagnetic fields. They are obtained from the Lindblad master equation,
\begin{equation}
\frac{\partial \hat{\rho}}{\partial t} =
-\frac{i}{\hbar}\,[\hat{H}',\hat{\rho}\,] + 
\frac{1}{2} \sum_{n} \left(
2\, \hat{C}_n \hat{\rho} \, \hat{C}_n^\dagger - \hat{C}_n^\dagger \hat{C}_n \hat{\rho} - \hat{\rho}\, \hat{C}_n^\dagger \hat{C}_n
\right),
\label{eq:Lindblad}
\end{equation}
where $\hat{\rho}$ is the density operator describing the state of the system and the $\hat{C}_n$'s are certain operators called jump (or collapse) operators. The latter
include the operator 
$\sqrt{\Gamma_{ij}} \, |\,i\,\rangle \langle\, j\,|$ if state $j$ relaxes to state $i$ at a rate $\Gamma_{ij}$ by spontaneous decay or some other mechanism. It is customary to add phenomenological terms in $-\gamma_{ij}\langle \, i\, |\,\hat{\rho}\, |\, j\,\rangle|\,i\,\rangle\langle\,j\,|$ and $-\gamma_{ij}\langle \, j\, |\,\hat{\rho}\, |\, i\,\rangle|\,j\,\rangle\langle\,i\,|$ to the right-hand side of Eq.~(\ref{eq:Lindblad}) if the coherences $\langle \, i\, |\,\hat{\rho}\, |\, j\,\rangle$  and $\langle \, j\, |\,\hat{\rho}\, |\, i\,\rangle$ decay at an additional rate $\gamma_{ij}$ due to pure dephasing effects such as collisional broadening.

The {\tt obe} and {\tt ldbl} modules calculate the density matrix, $\rho$, representing the density operator $\hat{\rho}$ in the $\{|i\rangle\}$ basis --- i.e., the elements of $\rho$ are the matrix elements of $\hat{\rho}$:
\begin{equation}
    \rho_{ij} = \langle\, i\, |\, \hat{\rho}\, | \, j\,\rangle, \qquad i,j = 1,\ldots,N.
\end{equation}
$\mbox{Re}\,\rho_{ij} = \mbox{Re}\,\rho_{ji}$ 
and
$\mbox{Im}\,\rho_{ij} = -\mbox{Im}\,\rho_{ji}$
since the density matrix is Hermitian.
These relations are used within the {\tt obe}, {\tt mbe} and {\tt ldbl} modules to store and calculate this matrix as a column vector of $N^2$ real numbers, ${\sf r}$, rather than as a 2D array of $N^2$ complex numbers.
Specifically, if the states are labelled 1, 2, 3,\ldots as done throughout this section,
\begin{equation}
{\sf r} =
\begin{pmatrix}
\rho_{11} & \mbox{Re}\, \rho_{12} & \mbox{Im}\, \rho_{12} &
\rho_{22} & \mbox{Re}\, \rho_{13} & \ldots &
\rho_{NN}
\end{pmatrix}^{\sf T}.
\label{eq:r_uppertr}
\end{equation}
Accordingly, the Lindblad equation is recast as a set of homogeneous linear relations between the elements of ${\sf r}$ and the elements of $\dot{\sf r}$, the time derivative of~${\sf r}$: 
\begin{equation}
\dot{\sf r} = {\sf L}\,{\sf r},
\label{eq:r_evol}
\end{equation}
where ${\sf L}$ is a $N^2 \times N^2$ real matrix. Much of the {\tt obe} and {\tt ldbl} code aim at constructing this matrix given the parameters of the system and at integrating Eq.~(\ref{eq:r_evol}), either as written or after further transformation. (Readers interested in knowing the details of how the matrix ${\sf L}$ is constructed are referred to the information given in the code of the subroutine
{\tt ldbl\_reformat\_rhs\_cmat} contained in
the subroutine {\tt ldbl\_set\_rhsmat} of the {\tt ldbl} module.) 


\subsubsection{Inhomogeneous broadening}
\label{section:Doppler}

The {\tt obe} and {\tt mbe} modules make it possible to take inhomogeneous broadening into account in the calculation of the density matrix. The codes are specifically geared towards the case of Doppler broadening arising from the free thermal motion of atoms in an atomic vapour. However, they can be easily generalised to other cases of Gaussian broadening if required. Extending them to Doppler broadening for a non-Maxwellian distribution of atomic velocities is also possible.

In the current state of development of the {\tt obe} and {\tt mbe} modules, Doppler averaging is possible only for co-propagating or counter-propagating co-linear fields. The internal state of an atom depends on the component of its velocity vector in the direction of propagation of the field, $v$, owing to the Doppler shift of the detunings $\Delta_\alpha$. To first order in $1/c$, 
\begin{equation}
    \Delta_\alpha(v) = \Delta_\alpha(v=0) - {k}_\alpha v
    \label{eq:positivev}
\end{equation}
if the wave vector ${\bf k}_\alpha$ is oriented in the positive $z$-direction or
\begin{equation}
    \Delta_\alpha(v) = \Delta_\alpha(v=0) + {k}_\alpha v
    \label{eq:negativev}
\end{equation}
if it is oriented in the negative $z$-direction. Correspondingly, the matrix ${\sf L}$ appearing in Eq.~(\ref{eq:r_evol}) depends on $v$, and so does the solution vector ${\sf r}$. Averaging the latter
over the Maxwellian distribution of atomic velocities gives the Doppler-averaged density matrix, $\rho^{\rm av}$, here represented by the column vector ${\sf r}^{\rm av}$:
\begin{equation}
{\sf r}^{\rm av} =
\int_{-\infty}^\infty {\sf r}(v)\, f_{\rm M}(v)\,{\rm d}v
\label{eq:rav}
\end{equation}
with
\begin{equation}
f_{\rm M}(v) = \frac{1}{u\sqrt{\pi}}\exp(-v^2/u^2).
\end{equation}
In this last equation, $u$ is the rms velocity of the atoms in the $z$-direction: $u = \sqrt{2k_{\rm B}T/M}$, where $k_{\rm B}$ is Boltzmann constant, $T$ is the temperature of the vapour and $M$ is the mass of the atom.

The {\tt obe} and {\tt mbe} modules include code calculating the integral over $v$ either by numerical quadrature or by expressing the integral in terms of the Faddeeva (or Faddeyeva) function, $w(z)$ \cite{Abramowitz}:
\begin{equation}
w(z) = \exp\left(-z^2\right) \left[1 + \frac{2{i}}{ \sqrt{\pi}}\int_0^z
\exp\left(t^2\right)\, {\rm d}t\, \right].
\end{equation}
In terms of the complementary error function \cite{Abramowitz},
\begin{equation}
w(z) = \exp\left(-z^2\right) \mbox{erfc}\,(-{i}z).
\end{equation}
The approach based on the Faddeeva function applies only to Doppler averaging of the steady state density matrix. It is outlined in \ref{appendix:steady} and \ref{appendix:single}.

The numerical quadrature method is more general. The
quadrature abscissas $\{v_k\}$ and quadrature weights $\{w_k\}$ used by the {\tt obe} and {\tt mbe} modules can either be provided by the user or calculated internally. As the program sets
\begin{equation}
    {\sf r}^{\rm av} = \sum_{k=1}^{N_v}
    w_k\,{\sf r}(v_k)\,f_{\rm M}(v_k),
    \label{eq:Dopplerquadrature}
\end{equation}
the quadrature weights should not include the velocity distribution $f_{\rm M}(v)$.
Since the Doppler effect is taken into account to first order in $1/c$ only, as per Eqs.~(\ref{eq:positivev}) and (\ref{eq:negativev}), the matrix ${\sf L}$ varies linearly with $v$:
\begin{equation}
{\sf L} = {\sf L}_0 + v\, {\sf L}_1,
\label{eq:Lv}
\end{equation}
where ${\sf L}_0$ and ${\sf L}_1$ do not depend on $v$. These two matrices are easily constructed, which makes Eq.~(\ref{eq:Lv}) an efficient way of re-calculating ${\sf L}$ for each value of $v$. Replacing $f_{\rm M}(v)$ by another velocity distribution, should this be necessary, would only require minor changes to the codes.


\subsubsection{Integrating the optical Bloch equations}
\label{section:timedep}

Integrating Eq.~(\ref{eq:r_evol}) subject to specified initial conditions gives the density matrix as a function of time.
Unless the size of the system is excessively large, this operation is amenable to standard numerical methods. This library provides five subroutines to this effect, namely {\tt obe\_Doppler\_av\_td\_A}, {\tt obe\_Doppler\_av\_td\_B} and
{\tt obe\_tdint}, for CW fields, and {\tt mbe\_tdint\_1} and {\tt mbe\_tdint\_2}, for fields with a time-dependent complex amplitude ${\cal E}_\alpha(t)$. The {\tt obe} routines can handle an arbitrary number of applied fields, whereas {\tt mbe\_tdint\_1} and {\tt mbe\_tdint\_2} are respectively limited to one and two fields. Both {\tt obe\_Doppler\_av\_td\_A} and {\tt obe\_Doppler\_av\_td\_B} calculate the Doppler-averaged time-dependent density matrix. These two routines differ by their memory and CPU times requirements. The {\tt obe\_tdint} routine calculates the time-dependent density matrix without Doppler averaging. Doppler averaging is optional for the two {\tt mbe} routines.

Each of these five routines offers a choice of integrator between the classic fourth-order Runge-Kutta method, Butcher's fifth-order Runge-Kutta method \cite{Butcher} and an adaptive ODE integrator (the DOP853 routine of Hairer et al, which is a Dormand-Prince implementation of an explicit eighth-order Runge-Kutta method~\cite{Hairer1993,dop853}).
A solution based on the right and left eigenvectors of the matrix ${\sf L}$ is also implemented, and
can be contemplated if this matrix
is time-independent (which is normally the case if the applied fields are CW). These eigenvectors fulfill the equations
\begin{equation}
{\sf L}{\sf v}_j = \lambda_j {\sf v}_j, \qquad j = 1,\ldots,{\cal N}
\end{equation}
and
\begin{equation}
{\sf u}_j^{\dagger} {\sf L} = \lambda_j {\sf u}_j^{\dagger}, \qquad j = 1,\ldots,{\cal N}
\end{equation}
with ${\cal N} = N^2$.
In many cases of interest, the initial density matrix vector can be written as a linear combination of the ${\sf v}_j$'s. I.e., there exist complex coefficients $c_1$, $c_2$, \ldots, $c_{\cal N}$ such that
\begin{equation}
    {\sf r}(t=t_0) = \sum_{j=1}^{\cal N} c_j \, {\sf v}_j.
\label{eq:expansion}
\end{equation}
In this case, the density matrix can be obtained for all times as
\begin{equation}
    {\sf r}(t) = \sum_{j=1}^{\cal N} c_j \,
    \exp[\lambda_j (t-t_0)]\,{\sf v}_j.
\label{eq:eigenmethod}
\end{equation}
However, the existence of such a set of coefficients is not guaranteed since the matrix
{\sf L} is not symmetric and may be defective. The subprogram {\tt obe\_tdint} offers the option to attempt to expand
${\sf r}(t=t_0)$ as per Eq.~(\ref{eq:expansion}) with
\begin{equation}
    c_j = \frac{{\sf u}_j^\dagger{\sf r}(t=t_0)}
    {{\sf u}_j^\dagger{\sf v}_j},
\end{equation}
and if this attempt is successful (it normally is), use Eq.~(\ref{eq:eigenmethod}) to propagate the density matrix in time.


\subsubsection{Rate equations}
\label{section:rate}

The optical Bloch equations can be transformed into
a smaller system of rate equations if the elements
of the density matrix
can be divided into two classes, ${\cal R}$ and ${\cal S}$, depending
on whether they converge to steady values
much more rapidly (${\cal R}$) or much more slowly
(${\cal S}$) than the elements belonging to the other class.
Class ${\cal R}$ typically includes
most or all the coherences, class ${\cal S}$
the populations and, if any, the coherences not included in ${\cal R}$.
This dichotomy makes it possible to reduce the number of coupled
differential equations
by adiabatic elimination of the elements belonging to ${\cal R}$. The details of this approach can be found in \ref{appendix:rate}.

The routines {\tt obe\_Doppler\_av\_td\_A} and {\tt obe\_tdint} can solve Eq.~(\ref{eq:r_evol}) within this approximation for a superposition of CW fields, with or without Doppler averaging. As the code is written, the set ${\cal S}$ of the elements of $\rho$ which are actually propagated in time includes all the populations and none of the coherences. The latter are derived from the former through Eq.~(\ref{eq:rR_evol2}) of \ref{appendix:rate}. Time propagation thus involves solving a system of only $N$ coupled differential equations, which is a considerable reduction from the original system of $N^2$ equations.


\subsubsection{Steady state solutions}
\label{section:steady}

In many cases, but not all cases, the populations and coherences
settle to constant values as time increases if the fields are CW. Then
${\sf r} \rightarrow {\sf r}_{\rm st}$ for $t \rightarrow \infty$, where
\begin{equation}
\dot{\sf r}_{\rm st} = {\sf L}\,{\sf r}_{\rm st} = 0.
\label{eq:r_st}
\end{equation}
The steady-state density matrix represented by the column vector
${\sf r}_{\rm st}$ is thus an eigenvector of the matrix ${\sf L}$
corresponding to a zero eigenvalue, and can usually be calculated
as such. The calculation follows the same lines as the calculation of ${\sf r}(t)$ by the eigenvalue method described in Section~\ref{section:timedep}, except that here only the eigenvectors ${\sf v}_j$ belonging to a zero eigenvalue are included in Eq.~(\ref{eq:eigenmethod}). The optical Bloch equations have no steady state solution if some of the eigenvalues $\lambda_j$ are imaginary.

The {\tt obe} module also supports a different way of obtaining the steady-state density matrix, which is based on transforming the eigenvalue equation ${\sf L}\,{\sf r} = 0$ into an inhomogeneous system of linear equations,
\begin{equation}
    {\sf L}'{\sf r}' = {\sf b},
    \label{eq:steadylinear}
\end{equation}
where ${\sf L}'$ is a $({\cal N}-1)\times ({\cal N}-1)$ square matrix and ${\sf b}$ is a $({\cal N}-1)$-component column vector. The matrix ${\sf L}'$ and the column vector ${\sf b}$ are derived from ${\sf L}$ by a straightforward rearrangement process. The transformation is normally possible due to the unit trace property of the density matrix, which constraints the solutions of this eigenvalue equation. The vector ${\sf r}_{\rm st}$ representing the steady state density matrix is identical to the solution vector ${\sf r}'$, apart from one population which can be calculated readily as a linear combination of the other populations. The reader is referred to \ref{appendix:steady} for the details of the method. 
Calculating ${\sf r}_{\rm st}$ in this way may be faster than by using the eigenvalue method but will fail if Eq.~(\ref{eq:r_st}) has more than one solutions. It would then be necessary to specify the density matrix that ${\sf r}_{\rm st}$ develops from in order to obtain a unique solution, which is not overly difficult in the eigenvalue method --- and is implemented in the {\tt obe} module --- but would considerably complicate the calculation based on the linear equations method.

Finding the steady state as per Eq.~(\ref{eq:steadylinear}) also makes it possible to Doppler average the density matrix semi-analytically, as an alternative on the entirely numerical approach mentioned in Section~\ref{section:Doppler}. 
This semi-analytical route may lead to substantial savings in CPU time as compared to a numerical quadrature.
Its principles are outlined in \ref{appendix:steady}.

As pointed out in that appendix, significant savings may also be achieved, along similar lines, in computations involving the calculation of the density matrix for multiple values of a same detuning.

Calculations of the steady-state density matrix are possible only for CW fields. Several routines are provided to this end, namely, for general systems, {\tt obe\_steadystate} (for calculations without Doppler averaging), {\tt obe\_Doppler\_av\_st} (for calculations with Doppler averaging performed semi-analytically as described in \ref{appendix:steady}, and {\tt obe\_Doppler\_av\_st\_numerical} (for calculations with Doppler averaging performed by a numerical quadrature). The library also includes a routine specialised to the case of 2-state systems driven by a single field ({\tt obe\_2state}), one specialised to the case of multi-state systems driven by a single field with the calculation organised as explained in \ref{appendix:steady} ({\tt obe\_steadystate\_onefld}), and, as described in next section, several routines specialised to calculations in the weak probe approximation.
{\tt obe\_steadystate} and {\tt obe\_Doppler\_av\_st\_numerical} can handle calculations using the eigenvalue method, which makes it possible to address cases for which the steady state depends on the initial populations. The subroutine {\tt obe\_Doppler\_av\_st\_numerical} is normally less efficient than {\tt obe\_Doppler\_av\_st}.


\subsubsection{The weak probe approximation}

These programs offer the option of solving the optical Bloch equations within the approximation where one of the fields is considered to be too weak to cause any appreciable optical pumping over the relevant time scales. 
A calculation within this approximation amounts to calculating the density matrix to first order in the weak field and to all orders in any of the other fields in the problem. The populations are not affected by the former in this case, while the coherences depend linearly on its amplitude, without any power broadening.

The weak field is referred to as the probe field in many applications of these methods, and the weak field approximation as the weak probe approximation. This terminology is also used here. How this approximation is implemented within the {\tt obe} module is explained in \ref{appendix:weak}.

The calculation of the steady state for a ladder system by the linear equations method may be problematic in the weak probe approximation. Ladder systems here refer to systems in which a set of low energy states, which are the only ones initially populated, are coupled to states of higher energy only by the probe field. The populations of the lower energy states do not vary in time in the weak probe approximation for such systems, and the populations of the higher energy states remain identically zero at all times. The steady state populations are thus the same as the initial ones, which are specified by the user. It is therefore possible to find the steady state coherences by an application of the rate equations method. Referring to \ref{appendix:rate}, the calculation simply amounts to solving Eq.~(\ref{eq:rR_evol2}) for the vector ${\sf r}_{\cal R}$, with ${\cal R}$ including all the coherences and ${\cal S}$ all the populations. Within this approach, folding the result on a Maxwellian distribution of atomic velocities can also be done in terms to the Faddeeva function, following the same method as outlined in 
\ref{appendix:steady}
but here starting from Eq.~(\ref{eq:rR_evol2}) rather than from Eq.~(\ref{eq:steadylinear}). 

Calculations within the weak probe approximation can normally be handled by the general computational routines contained in the {\tt obe} module. However, steady state calculations for ladder system are best done by the subroutine {\tt obe\_steadystate\_ladder}. Specialised subroutines ({\tt obe\_weakprb\_3stladder} and {\tt obe\_weakprb\_4stladder}) are also provided for calculations in the weak probe approximation for, respectively, 3-state ladder systems \cite{Gea-Banacloche1995,Tanasittikosol2011} and 4-state ladder systems \cite{4state}.

The steady state density matrix takes on a particularly simple form in the weak probe approximation if the system comprises only two states or two groups of states coupled by a single field. This case is described in \ref{appendix:single}. Three specialised routines are provided for tackling such systems, namely {\tt obe\_steadystate\_onefld\_weakprb}, {\tt obe\_weakfield} (a stand-alone subprogram which also calculates the complex susceptibility, the refractive index and the absorption coefficient), and 
{\tt obe\_steadystate\_onefld\_powerbr}
(for systems in which power broadening is significant but optical pumping is not, as explained in the detailed description of this subprogram given in the User Manual).


\subsubsection{The complex susceptibility}
\label{section:chi}

Let ${\bf P}(t)$ be the polarisation generated in the medium by the optical field described by Eq.~(\ref{eq:Etot}). (As mentioned above, we set ${\bf r} = 0$ in this equation. The $\exp(\pm i\,{\bf k}_\alpha\cdot{\bf r})$ phase factors are assumed to be taken into account through the complex amplitudes ${\cal E}_\alpha$ if they are relevant.) 
In terms of a complex susceptibility $\chi(\omega_\alpha)$,
\begin{align}
{\bf P}(t) &= \frac{\epsilon_0}{2}\,\sum_{\alpha = 1}^M \hat{\text{\boldmath{$\epsilon$}}}_\alpha \, {\cal E}_\alpha \chi(\omega_\alpha)
\exp(-i\omega_\alpha t) + \mbox{c.c.} + \ldots,
\end{align}
where the $\ldots$ stand for contributions oscillating at frequencies other than $\omega_\alpha$, if any is present.
It is assumed, in the following, that these additional contributions are negligible or absent. 

For such systems,
\begin{equation}
    \chi(\omega_\alpha) = 2 N_{\rm d} {\sum_{i,j}}^\prime\, \rho_{ij}\,
    \langle\, j\, |\,
\hat{\text{\boldmath{$\epsilon$}}}^*_\alpha \cdot \hat{\bf D} \,|\, i\, \rangle /(\epsilon_0 \, {\cal E}_\alpha),
\label{eq:chifirst}
\end{equation}
where $N_{\rm d}$ is the number density
and, as in Section~\ref{section:generalformulation}, $\hat{\bf D}$ is the dipole operator.
For each field, the summation runs over all the states $|i\rangle$ and all the states $|j\rangle$ dipole-coupled to each other by this field and such that $\hbar\omega^{(i)} > \hbar\omega^{(j)}$ with $\omega^{(i)} - \omega^{(j)} \approx \omega_\alpha$.
This equation can also be written in the following form, which is the one implemented in the programs:
\begin{equation}
    \chi(\omega_\alpha) = 2 N_{\rm d} {\sum_{i,j}}^\prime\, \rho_{ij}\,
    \langle\, i\, |\,
\hat{\text{\boldmath{$\epsilon$}}}_\alpha \cdot \hat{\bf D} \,|\, j\, \rangle^* /(\epsilon_0 \, {\cal E}_\alpha).
\label{eq:chi}
\end{equation}
The coherences $\rho_{ij}$'s and therefore the susceptibility $\chi(\omega_\alpha)$ generally depend on the intensity of all the fields included in the calculation --- with the important exception of systems containing only one field and this field is treated within the weak probe approximation (see \ref{appendix:single}).


Besides the complex susceptibility, the programs can also calculate the
corresponding refractive index, $n(\omega_\alpha)$, and absorption coefficient, $\alpha(\omega_\alpha)$. Here \cite{Loudon}, 
\begin{align}
    n(\omega_\alpha) &= \mbox{Re}\,[1 + \chi(\omega_\alpha)]^{1/2}\\
    \alpha(\omega_\alpha) &= 2 k_\alpha \,\mbox{Im}\,[1 + \chi(\omega_\alpha)]^{1/2}.
\end{align}

The library contains two routines calculating these quantities, namely
{\tt obe\_susceptibility}, which uses pre-calculated coherences, and {\tt obe\_weakfield}, which is self-contained and computes the necessary coherences within the weak probe approximation for multi-state single-field systems. 


\subsection{The Maxwell-Bloch equations}


\subsubsection{General formulation}

The {\tt mbe} module addresses the case of a single field or a superposition of two different fields, i.e., a probe field and a coupling field, (co)propagating in the positive $z$-direction. Solving the Maxwell-Bloch equations for more than two fields or in another geometry is not yet supported.

In general, the spatial and temporal variation of the electric field component of the electromagnetic wave is governed by the equation
\begin{equation}
\nabla^2 {\bf E} - \frac{1}{c^2}
\frac{\partial^2 {\bf E}}{\partial t^2} = \mu_0 \,\frac{\partial^2 {\bf P}}{\partial t^2},
\end{equation}
where ${\bf P}$ is the medium polarisation and $\mu_0$ is the vacuum permeability. The plane wave approximation is assumed in the calculation performed by the {\tt mbe} codes. I.e., it is assumed that ${\bf E}$ and ${\bf P}$ are constant in any plane perpendicular to the $z$-axis. These fields thus depend only on $z$ and $t$, and the 3D wave equation reduces to the 1D equation
\begin{equation}
\frac{\partial^2 {\bf E}}{\partial z^2} - \frac{1}{c^2}
\frac{\partial^2 {\bf E}}{\partial t^2} = \mu_0 \,\frac{\partial^2 {\bf P}}{\partial t^2}.
\end{equation}
This equation can be simplified further, to
\begin{equation}
\frac{\partial {\cal E}_\alpha}{\partial z} + \frac{1}{c}
\frac{\partial {\cal E}_\alpha}{\partial t} = i\,\frac{k_\alpha}{2\epsilon_0}\,{\cal P}_\alpha(z,t),
\label{eq:propafirst}
\end{equation}
by making the ansatz 
\begin{align}
   {\bf E}(z,t) &= \frac{1}{2}\,\sum_{\alpha=1}^M \, \hat{\text{\boldmath{$\epsilon$}}}_\alpha \, {\cal E}_\alpha(z,t)
\exp[i({\bf k}_\alpha\cdot{\bf r} - \omega_\alpha t)] + \mbox{c.c.}, \\
{\bf P}(z,t) &= \frac{1}{2}\,\sum_{\alpha=1}^M \, \hat{\text{\boldmath{$\epsilon$}}}_\alpha \, {\cal P}_\alpha(z,t)
\exp[i({\bf k}_\alpha\cdot{\bf r} - \omega_\alpha t)] + \mbox{c.c.},
\end{align}
and taking into account that the complex amplitudes ${\cal E}_\alpha(z,t)$ and ${\cal P}_\alpha(z,t)$ vary slowly compared to the carriers. As noted above, the library only supports calculations for a single field ($M = 1$) or a superposition of two fields ($M = 2$). The field with $\alpha=1$ is referred to as the probe field and the field with $\alpha=2$ (if present) as the coupling field.

The relationship between the medium polarisation and the state of the atoms is considered in Section~\ref{section:chi}, from which it follows that
\begin{equation}
    {\cal P}_\alpha(z,t) = 2 N_{\rm d} {\sum_{i,j}}^\prime\, \rho_{ij}\,
    \langle\, i\, |\,
\hat{\text{\boldmath{$\epsilon$}}}_\alpha \cdot \hat{\bf D} \,|\, j\, \rangle^*,
\end{equation}
where $N_{\rm d}$ is the medium number density and the summation runs as in Eqs.~(\ref{eq:chifirst}) and (\ref{eq:chi}).
Changing the time variable $t$ to the shifted time $t'$, with
\begin{equation}
    t' = t - z/c,
\end{equation}
further simplifies
Eq.~(\ref{eq:propafirst}) to 
\begin{equation}
\frac{\partial {\cal E}_\alpha}{\partial z} = i\,\frac{N_{\rm d}k_\alpha}{\epsilon_0}\,{\sum_{i,j}}^\prime\, \rho_{ij}\,
    \langle\, i\, |\,
\hat{\text{\boldmath{$\epsilon$}}}_\alpha \cdot \hat{\bf D} \,|\, j\, \rangle^*,
\label{eq:propa}
\end{equation}
where ${\cal E}_\alpha$ and the coherences $\rho_{ij}$ are now functions of $z$ and $t'$ rather than functions of $z$ and $t$.
This last equation governs the propagation of the fields through the medium, as calculated by the present programs.

\subsubsection{Implementation}
\label{section:mbeimplementation}
 
The subroutines {\tt mbe\_propagate\_1} and {\tt mbe\_propagate\_2} solve Eq.~(\ref{eq:propa}), with the coherences obtained by solving Eq.~(\ref{eq:Lindblad}), respectively for the case of a single field ($\alpha = 1$) or a superposition of two fields ($\alpha = 1,2$).
The calculation yields the density matrix describing the state of the medium, $\rho(z,t')$, and the complex amplitude(s) of the propagated field(s), ${\cal E}_{\alpha}(z,t')$. These results are calculated on a two-dimensional mesh of values of $z$ and $t'$. The grid points in the $z$-direction extend from $z=z_0 = 0$ (the entrance of the medium) to $z=z_{\rm max}$ and are separated by a constant step $h$:
$$z = z_i = z_0 + ih, \quad i = 0,\ldots,N_z,$$
with $z_0 = 0$ and $h = z_{\rm max}/N_z$. The distance $z_{\rm max}$ and the number of spatial steps, $N_z$, are set by the user. 
The complex amplitude of the fields at $z_0$ must be provided on a mesh of $N_t+1$ values of $t$, namely at $t=t_k$ with
$k=0$, 1,\ldots, $N_t$. The same mesh is used by {\tt mbe\_propagate\_1} and {\tt mbe\_propagate\_2} for the shifted time $t'$. Namely, at all z, the grid points in the $t'$-direction are taken to be at 
\begin{displaymath}
t' = t'_{k} = t_k, \quad k = 0,\ldots,N_t.
\end{displaymath}

The calculation alternates at each spatial step between obtaining the coherences $\rho_{ij}(z,t')$ given the field(s) and propagating the field(s) to the next step given these coherences. If Doppler averaging is required, the coherences are obtained for a number of velocity classes and their average, weighted by the Maxwellian velocity distribution, is calculated by numerical quadrature.

The density matrix is calculated at each $z_i$ by integrating the optical Bloch equations, starting, at $t' = t'_{0}$, with initial values determined by the user. A fourth order Runge Kutta rule is used to this end for the integration in time and a predictor-corrector method combining the third order Adams-Bashford rule and the fourth order Adams-Moulton rule for the integration in space. Other choices of methods are also offered.
More information about the different possibilities can be found in the User Manual.



\section{Using this software}
\label{section:using_honours}

\subsection{Installation}
\label{section:installation}

The most recent version of these modules can be found at the URL  https://github.com/durham-qlm/CoOMBE.
Installing this software only involves downloading the {\tt general\_settings.f90}, {\tt obe.f90}, {\tt mbe.f90}, {\tt ldbl.f90} and {\tt driveall.f90} files, and editing the {\tt general\_settings.f90} file as required (see Section~\ref{section:keyparameters}). The latter is the only program file which may need customisation.

A calculation using these modules requires a driving program, which could be either the {\tt driveall} program provided in the {\tt driveall.f90} file or a user-written bespoke Fortran~90 program. The {\tt driveall} program is described in Section~\ref{section:driveall}. Information relevant for the development of a bespoke driver can be found in Section~\ref{section:bespoke}. The User Manual included in this distribution contains  detailed information about the use of {\tt driveall} and (for Fortran programmers) the use of the various user-facing subroutines contained in these modules.

Running this software requires compiling the programs and linking it to the LAPACK \cite{Lapack} and BLAS \cite{Blas} libraries.
If a Fortran compiler and these two libraries are already installed,
compiling these programs could be done, e.g., by the command\footnote{This command is given for illustrative purpose only. How to invoke the compiler and link the LAPACK and BLAS libraries is system dependent and may vary from installation to installation.}
\begin{lstlisting}
gfortran general_settings.f90 ldbl.f90 obe.f90               mbe.f90 driveall.f90 -llapack -lblas 
\end{lstlisting}
and similarly for a bespoke program. In the latter case, the {\tt mbe.f90} file does not need to be compiled if the program does not call any of the {\tt mbe} subroutine listed in Table~\ref{table:obembe_contents}.

We also provide advice, in \ref{appendix:podman}, for compiling and running these programs through a container, specifically a Podman container \cite{podman}. This alternative, self-contained method allows the software to be used without installing a Fortran compiler or any supporting libraries directly to the user's machine. This feature is offered for convenience to users not familiar with compiling Fortran codes and does not limit the scope of the program or its output. The distribution includes the files required for running all the examples provided in the {\tt examples} directory in this way.


\subsection{Key parameters}
\label{section:keyparameters}

The following parameters are defined in the {\tt general\_settings} module and must be adapted to the requirements of the intended calculation before compilation:
\begin{description}[align=left,labelwidth=0.9cm,leftmargin=1.1cm]
\item[$\;\;${\tt nst}:] 
An {\tt integer} constant which must be given a value equal to the number of states in the model, ${N}$. Changing the value of {\tt nst} is the only editing which may be required across all the modules for adapting the Fortran code to the problem at hands.
\item[$\;\;${\tt kd}:] An {\tt integer} defining the {\tt kind} of many of the variables used in {\tt obe} and {\tt mbe} --- i.e., defining whether these variables are of {\tt real} or {\tt double precision} type ({\tt complex} or {\tt double complex} for variables storing complex numbers). Selecting a {\tt kind} parameter corresponding to {\tt real} variables rather than to {\tt double precision} variables will reduce memory requirements and computation time but may also result in larger numerical inaccuracies.
\item[$\;\;${\tt nmn}:] An {\tt integer} constant defining how the states are numbered by the user, as explained in Section~\ref{section:numbering}.
\end{description}


\subsection{Required data}
\label{section:data}

The routines provided require various input data, which are problem-dependent and need to be prepared separately. These will typically include:
\begin{itemize}
    \item the energy offset $\hbar\delta\omega^{(i)}$ defined by Eq.~(\ref{eq:offsets}) for each of the states considered;
\item the rates of spontaneous decay $\Gamma_{ij}$ from a state $j$ to a state $i$, for all the states considered;
\item any additional dephasing rate $\gamma_{ij}$ that would need to be included in the calculation to take into account the frequency widths of the fields and/or other pure dephasing effects;
\item the detuning $\Delta_{\alpha}$, complex field amplitude ${\cal E}_\alpha$ and wavelength of each of the fields considered, as well as the transition dipole moments for each of the transitions driven by these fields or the corresponding Rabi frequencies;
\item the initial populations (i.e., the initial values of the diagonal elements of the density matrix);
\item the temporal profile of the applied field(s), unless these fields are CW or their profile can be calculated internally;
\item the frequency widths of the fields, if these widths should be taken into account otherwise than through the rates $\gamma_{ij};$
\item the atomic number and the wavelength of each field density in the case of a propagation calculation or a calculation of the complex susceptibility.
\end{itemize}
All energies and angular frequencies are to be provided as frequencies specified in MHz. E.g., the energy offset $\hbar\delta \omega^{(i)}$ needs to be provided as the frequency $\delta \omega^{(i)}/(2\pi)$. Wavelengths are to be expressed in nm, densities in number of atoms per m$^3$, dipole moments in C~m and electric field amplitudes in V~m$^{-1}$.

Besides the wavelength of each field, calculations involving Doppler averaging will also require the r.m.s.\ thermal speed of the atoms in the laser propagation direction, $u_{}$, in m~s$^{-1}$, and the abscissas and integration weights for the numerical quadrature over the atoms' velocity distribution (unless the calculation uses one of the quadrature rules provided by {\tt obe} or the integration is done analytically using the Faddeeva function).


\subsection{Running these codes through the {\tt driveall} program}
\label{section:driveall}


\begin{table}[t]
\caption{\label{table:keyparams}Contents of the keyparams file read by the {\tt driveall} program. Comprehensive information about these parameters and the use of {\tt driveall} can be found in the User Manual.}
\centering
\footnotesize
\begin{tabular}{l l}
{}\\[-2mm]
\hline\\ { }\\[-6mm] 
\parbox{3.1cm}{Name} & Short description\\
{ }\\[-3mm]
\hline\\
{}\\[-4mm]
\multicolumn{2}{l}{\underline{Mandatory parameters}} \\
{}\\[-2mm]
{\tt nfields}&\parbox[t]{4.0cm}{
The number of fields.
}\\[-3mm]
\\
{\tt nstates}&\parbox[t]{4.0cm}{
The number of states.
}\\[-3mm]
\\
{\tt nmin}&\parbox[t]{4.0cm}{
The starting number in the indexing of the states.
}\\[-3mm]
\\
{\tt filename\_controlparams}&\parbox[t]{4.0cm}{
The name of the {\tt controlparams} file.
}\\[-3mm]
%
\\[3mm]
\multicolumn{2}{l}{\underline{Optional parameters}} \\
{}\\[-2mm]
{\tt filename\_defaultdata}&\parbox[t]{4.0cm}{
The name of the {\tt defaultdata} file.
}\\[-3mm]
\\
{\tt icmplxfld}&\parbox[t]{4.0cm}{
Parameter indicating whether the field amplitudes, dipole moments and Rabi frequencies are specified as real numbers or as complex numbers in the input files.
}\\[-3mm]
\\
{ }\\[-2mm]
\hline 
\end{tabular}
\end{table}

\begin{table*}[t]
\caption{\label{table:controlparams}Contents of the controlparams file read by the {\tt driveall} program: I. General parameters. Comprehensive information about these parameters and the use of {\tt driveall} can be found in the User Manual. The parameters indicated by an asterisk may be specified in the {\tt defaultdata} file rather than in the {\tt controlparams} file.}
\centering
\footnotesize
\begin{tabular}{l l}
{}\\[-2mm]
\hline\\ { }\\[-6mm] 
\parbox{4.6cm}{Name} & Short description\\
{ }\\[-3mm]
\hline\\
{}\\[-4mm]
\multicolumn{2}{l}{\underline{Mandatory parameters}} \\
{}\\[-2mm]
 {\tt icalc}&\parbox[t]{12.7cm}{
 Parameter determining whether the program should calculate the steady state density matrix, or the density matrix as a function of time, or integrate the Maxwell-Bloch equations.
}\\[-3mm]
\\
 {\tt iRabi}&\parbox[t]{12.7cm}{
 Parameter determining whether the relevant Rabi frequencies are input data or are to be calculated by the program.
}\\[-3mm]

\\[3mm]
\multicolumn{2}{l}{\underline{Optional parameters}} \\
{}\\[-2mm]
 (*) {\tt add\_dephas}&\parbox[t]{12.7cm}{
The dephasing rates $\gamma_{ij}$ divided by $2\pi$, in MHz, if relevant.
}\\[-3mm]
\\
 (*) {\tt amplitude}&\parbox[t]{12.7cm}{
 The electric field amplitudes ${\cal E}_\alpha$ in V~m$^{-1}$, expressed as real numbers.
}\\[-3mm]
\\
 (*) {\tt camplitude}&\parbox[t]{12.7cm}{
 The electric field amplitudes ${\cal E}_\alpha$ in V~m$^{-1}$, expressed as complex numbers.
 }\\[-3mm]
\\
 (*) {\tt cdip\_mom}&\parbox[t]{12.7cm}{
 The electric dipole moments $\langle\, i\, |\,
\hat{\text{\boldmath{$\epsilon$}}}_{\alpha} \cdot \hat{\bf D} \,|\, j\, \rangle$ in C~m, expressed as complex numbers. 
 }\\[-3mm]
\\
 (*) {\tt cRabif}&\parbox[t]{12.7cm}{
 The Rabi frequencies $\Omega_{\alpha;ij}$ divided by $2\pi$, in MHz, expressed as complex numbers. 
 }\\[-3mm]
\\
 (*) {\tt detuning}&\parbox[t]{12.7cm}{
 The detunings $\Delta_\alpha$ divided by $2\pi$, in MHz.
 }\\[-3mm]
\\
 (*) {\tt detuning\_fact}&\parbox[t]{12.7cm}{
 The detuning factors $a_{i\alpha}$.
 }\\[-3mm]
\\
 (*) {\tt dip\_mom}&\parbox[t]{12.7cm}{
  The electric dipole moments $\langle\, i\, |\,
\hat{\text{\boldmath{$\epsilon$}}}_{\alpha} \cdot \hat{\bf D} \,|\, j\, \rangle$ in C~m, expressed as real numbers.
 }\\[-3mm]
\\
 (*) {\tt energ\_f}&\parbox[t]{12.7cm}{
 The energy offsets $\delta \omega^{(i)}$ divided by $\hbar$, in MHz.
 }\\[-3mm]
\\
 {\tt filename\_rho\_out}
 &\parbox[t]{12.7cm}{
 The name(s) of the output file(s) to which the program should write the density matrix.
 }\\[-3mm]
\\
 (*) {\tt Gamma\_decay\_f}&\parbox[t]{12.7cm}{
 The spontaneous decay rates $\Gamma_{ij}$ divided by $2\pi$, in MHz.
 }\\[-3mm]
\\
{\tt iappend}&\parbox[t]{12.7cm}{
Parameter determining whether existing output files can be overwritten with new results. 
}\\[-3mm]
\\
 {\tt iDoppler}&\parbox[t]{12.7cm}{
 Parameter determining whether the density matrix must be Doppler-averaged.
 }\\[-3mm]
\\
 {\tt iweakprb}&\parbox[t]{12.7cm}{
 Parameter determining whether the calculation is to be done within the weak probe approximation.
 }\\[-3mm]
\\
 (*) {\tt Rabif}&\parbox[t]{12.7cm}{
  The Rabi frequencies $\Omega_{\alpha;ij}$ divided by $2\pi$, in MHz, expressed as real numbers.
 }\\[-3mm]


\\
{ }\\[-2mm]
\hline 
\end{tabular}
\end{table*}
\begin{table*}[t]
\caption{\label{table:controlparams_123}Contents of the controlparams file read by the {\tt driveall} program: II. Parameters specific to particular types of calculation. Comprehensive information about these parameters and the use of {\tt driveall} can be found in the User Manual. The parameters indicated by an asterisk may be specified in the {\tt defaultdata} file rather than in the {\tt controlparams} file.}
\centering
\footnotesize
\begin{tabular}{l l}
{}\\[-2mm]
\hline\\ { }\\[-6mm] 
\parbox{4.6cm}{Name} & Short description\\
{ }\\[-3mm]
\hline\\
{}\\[-4mm]
\multicolumn{2}{l}{\underline{Parameters to be provided for time-dependent calculations of the density matrix}} \\
{}\\[-2mm]
 {\tt imethod}&\parbox[t]{12.7cm}{
Parameter determining which numerical algorithm is to be used in the integration of the optical Bloch equation.
 }\\[-3mm]
\\
 {\tt n\_time\_steps}&\parbox[t]{12.7cm}{
 $N_t$, the number of integration steps in the time integration of the optical Bloch equations.
 }\\[-3mm]
\\
 (*) {\tt popinit}&\parbox[t]{12.7cm}{
 The initial populations.
 }\\[-3mm]

\\[3mm]
\multicolumn{2}{l}{\underline{Optional parameters relevant for time-dependent calculations of the density matrix}} \\
{}\\[-2mm]
 {\tt filename\_rhoall\_out}&\parbox[t]{12.7cm}{
 The name of the file to which the whole density should be written at each time step.
 }\\[-3mm]
\\
 {\tt iAorB}&\parbox[t]{12.7cm}{
 Parameter determining which subroutine should be used in calculations with Doppler averaging.
 }\\[-3mm] 
\\
 {\tt inoncw}&\parbox[t]{12.7cm}{
 Parameter determining whether the calculation is for CW fields or for fields with a time-dependent envelope.
 }\\[-3mm]
\\
 {\tt iprintrho}&\parbox[t]{12.7cm}{
 Parameter determining whether the final density matrix is to be written out to the standard output stream.
 }\\[-3mm] 
\\
 {\tt irate}&\parbox[t]{12.7cm}{
 Parameter determining whether the calculation is to be done within the rate equations approximation.
 }\\[-3mm]
\\
 {\tt iunformatted}&\parbox[t]{12.7cm}{
 Parameter determining whether the output files should be unformatted (binary) rather than formatted.
 }\\[-3mm]
\\
 {\tt ti}, {\tt tf}&\parbox[t]{12.7cm}{
 The lower and upper bounds of the time integration interval, in $\mu$s. 
 }\\[-3mm]

\\[3mm]
\multicolumn{2}{l}{\underline{Optional parameters relevant for calculations of the steady state density matrix}} \\
{}\\[-2mm]
{\tt filename\_chi\_out}&\parbox[t]{12.7cm}{
The name of the file to which the calculated complex susceptibilities, refractive indexes and absorption coefficients should be written.
 }\\[-3mm]
\\
 {\tt iDoppler\_numer\_st}&\parbox[t]{12.7cm}{
 Parameter determining whether Doppler averaging is to be done analytically or numerically.
 }\\[-3mm]
\\
 {\tt iladder\_wkprb}&\parbox[t]{12.7cm}{
 Parameter determining whether the steady state density matrix is to be calculated by a subroutine specialised to ladder systems rather than by general subroutines.
 }\\[-3mm]
\\
 {\tt ioption}&\parbox[t]{12.7cm}{
 Parameter determining the algorithm used for calculating the steady state density matrix. 
 }\\[-3mm]
\\
 {\tt iprintrho}&\parbox[t]{12.7cm}{
 Parameter determining whether the steady state density matrix is to be written out to the standard output stream. 
 }\\[-3mm]
\\
 {\tt isuscept}&\parbox[t]{12.7cm}{
 Parameter determining whether the complex susceptibility at the probe frequency and the corresponding values of the refractive index and absorption coefficient are calculated after the steady state density matrix has been obtained. 
 }\\[-3mm] 
\\
 {\tt ivarydetuning}&\parbox[t]{12.7cm}{
 Parameter determining whether the steady state density matrix is to be calculated over a range of detunings.
 }\\[-3mm]
\\
 (*) {\tt popinit}&\parbox[t]{12.7cm}{
  The initial populations.
 }\\[-3mm]
\\[3mm]
\multicolumn{2}{l}{\underline{Parameters to be provided in the case of propagation calculations}} \\
{}\\[-2mm]
 (*) {\tt density}&\parbox[t]{12.7cm}{
 The density of the medium expressed as the number of atoms per m$^3$.
 }\\[-3mm]
\\
 {\tt imethod}&\parbox[t]{12.7cm}{
 Parameter determining which numerical method is to be used for integrating the optical Bloch equations.
 }\\[-3mm]
\\
 {\tt n\_time\_steps}&\parbox[t]{12.7cm}{
 $N_t$, the number of integration steps in the time integration of the optical Bloch equations.
 }\\[-3mm]
\\
 {\tt n\_z\_steps}&\parbox[t]{12.7cm}{
 The number of integration steps to be taken between $z=0$ and $z=z_{\rm max}$.
 }\\[-3mm]
\\
 (*) {\tt popinit}&\parbox[t]{12.7cm}{
  The initial populations.
 }\\[-3mm]
 \\
{\tt wavelength}&\parbox[t]{12.7cm}{
The wavelength(s) of the field(s) considered in nm.
}\\[-3mm]
\\
 {\tt zmax}&\parbox[t]{12.7cm}{
The distance over which the field(s) must be propagated, $z_{\rm max}$, in $\mu$m.
 }\\[-3mm]

\\[3mm]
\multicolumn{2}{l}{\underline{Optional parameters relevant for propagation calculations}} \\
{}\\[-2mm]
 {\tt filename\_rhoall\_out}&\parbox[t]{12.7cm}{
The name of the file to which the whole density matrix should be written. 
 }\\[-3mm]
\\
 {\tt iunformatted}&\parbox[t]{12.7cm}{
  Parameter determining whether the output files should be unformatted (binary) rather than formatted.
 }\\[-3mm]
\\
 {\tt izrule}&\parbox[t]{12.7cm}{
 Parameter determining the numerical algorithm used in the spatial propagation.
 }\\[-3mm] 
\\
 {\tt nt\_writeout}, {\tt nz\_writeout}&\parbox[t]{12.7cm}{
 Constants determining at which time or spatial steps results should be written out.
 }\\[-3mm]
\\
 {\tt ti}, {\tt tf}&\parbox[t]{12.7cm}{
 The lower and upper bounds of the time integration interval, in $\mu$s. 
 }\\[-3mm]

\\
{ }\\[-2mm]
\hline 
\end{tabular}
\end{table*}
\begin{table*}[t]
\caption{\label{table:controlparams_misc}Contents of the controlparams file read by the {\tt driveall} program: III. Miscellaneous parameters. Comprehensive information about these parameters and the use of {\tt driveall} can be found in the User Manual. The parameters indicated by an asterisk may be specified in the {\tt defaultdata} file rather than in the {\tt controlparams} file.}
\centering
\footnotesize
\begin{tabular}{l l}
{}\\[-2mm]
\hline\\ { }\\[-6mm] 
\parbox{4.6cm}{Name} & Short description\\
{ }\\[-3mm]
\hline\\
{}\\[-4mm]
\multicolumn{2}{l}{\underline{Parameters to be provided for calculations with analytical Doppler-averaging}} \\
{}\\[-2mm]

 {\tt idir}&\parbox[t]{12.7cm}{
 Parameter(s) indicating whether the corresponding fields propagate in the positive $z$-direction or the negative $z$-direction.
 }\\[-3mm]  
\\
 (*) {\tt urms}&\parbox[t]{12.7cm}{
 The root-mean squared velocity of the atoms in the laser propagation direction, $u$, in m~s$^{-1}$.
 }\\[-3mm]
\\
 (*) {\tt wavelength}&\parbox[t]{12.7cm}{
 The wavelength(s) of the field(s) considered, in nm.
 }\\[-3mm]

\\[3mm]
\multicolumn{2}{l}{\underline{Parameters to be provided for calculations with numerical Doppler-averaging}} \\
{}\\[-2mm]
 {\tt filename\_Dopplerquad}&\parbox[t]{12.7cm}{
 The name of the file containing the quadrature abscissas and quadrature weights to be used in the calculation, if this file is required.
 }\\[-3mm]
\\
 {\tt idir}&\parbox[t]{12.7cm}{
  Parameter(s) indicating whether the corresponding fields propagate in the positive $z$-direction or the negative $z$-direction.
 }\\[-3mm]
\\
 {\tt irule}&\parbox[t]{12.7cm}{
 Parameter determining which quadrature rule is to be used in the calculation.
 }\\[-3mm]
\\
 {\tt n\_v\_values}&\parbox[t]{12.7cm}{
The number of integration points in the integration over $v$.
 }\\[-3mm]
\\
 (*) {\tt urms}&\parbox[t]{12.7cm}{
  The root-mean squared velocity of the atoms in the laser propagation direction, $u$, in m~s$^{-1}$.
 }\\[-3mm]
\\
 {\tt vmax}&\parbox[t]{12.7cm}{
 The maximum value of $|v|$ to be considered, in m~s$^{-1}$.
 }\\[-3mm]
\\
 (*) {\tt wavelength}&\parbox[t]{12.7cm}{
 The wavelength(s) of the field(s) considered, in nm.
 }\\[-3mm]

\\[3mm]
\multicolumn{2}{l}{\underline{Parameters to be provided when the DOP853 ODE integrator is to be used}} \\
{}\\[-2mm]
 {\tt atol}&\parbox[t]{12.7cm}{
 Parameter controlling the allowed absolute error on the populations and coherences calculated by the program.
 }\\[-3mm]
\\
 {\tt rtol}&\parbox[t]{12.7cm}{
 Parameter controlling the allowed relative error on the populations and coherences calculated by the program. 
 }\\[-3mm]

\\[3mm]
\multicolumn{2}{l}{\underline{Parameters relevant for calculations involving non-CW fields}} \\
{}\\[-2mm]
 {\tt filename\_tdamps\_in}&\parbox[t]{12.7cm}{
The name of the file containing the time-dependent amplitude(s) of the field(s) considered, if this file is required. 
 }\\[-3mm] 
\\
 {\tt filename\_tdamps\_out}&\parbox[t]{12.7cm}{
 The name of the file to which the program should write the time-dependent amplitude(s) of the field(s) considered.
 }\\[-3mm]
\\
 {\tt iforce0}&\parbox[t]{12.7cm}{
 Parameter(s) determining whether the corresponding field should be taken to be initially zero.
 }\\[-3mm]
\\
 {\tt iinterp}&\parbox[t]{12.7cm}{
 Parameter determining whether tabulated field amplitudes should be interpolated.
 }\\[-3mm] 
\\
 {\tt istart}&\parbox[t]{12.7cm}{
 Parameter determining the initial values of the populations and coherences.
 }\\[-3mm]
\\
 {\tt itdfieldsAorB}&\parbox[t]{12.7cm}{
 Parameter determining whether the amplitude(s) of the applied field(s) must be calculated by the program or read from file.
 }\\[-3mm]
\\
 {\tt nsubsteps}&\parbox[t]{12.7cm}{
 Number of substeps within each integration step.
 }\\[-3mm]
\\
 {\tt pulse\_type}&\parbox[t]{12.7cm}{
 Parameter(s) determining the pulse envelope of the respective field.
 }\\[-3mm]
\\
 {\tt t0}, {\tt t1}, {\tt tw}&\parbox[t]{12.7cm}{
 Parameters determining the shape of the pulse envelope of the respective field.
 }\\[-3mm]

\\[3mm]
\multicolumn{2}{l}{\underline{Parameters to be provided for steady state calculations over a range of detunings}} \\
{}\\[-2mm]

 {\tt detuning\_min}, {\tt detuning\_max}&\parbox[t]{12.7cm}{
 The smallest and largest values of the detuning divided by $2\pi$, in MHz.
 }\\[-3mm]
 \\
  {\tt detuning\_step}&\parbox[t]{12.7cm}{
  The step in detuning, divided by $2\pi$, in MHz.
 }\\[-3mm]
\\
 {\tt index\_field}&\parbox[t]{12.7cm}{
 Parameter determining which field should be varied in the calculation.
 }\\[-3mm]

\\[3mm]
\multicolumn{2}{l}{\underline{Parameters to be provided for a calculation of the susceptibility at the probe frequency}} \\
{}\\[-2mm]

 (*) {\tt density}&\parbox[t]{12.7cm}{
  The density of the medium expressed as the number of atoms per m$^3$.
 }\\[-3mm]
\\
 (*) {\tt wavelength}&\parbox[t]{12.7cm}{
 The wavelength of the probe field in nm.
 }\\[-3mm]  


\\
{ }\\[-2mm]
\hline 
\end{tabular}
\end{table*}

All the features of the {\tt obe} and {\tt mbe} routines are accessible through the {\tt driveall} program, with the exceptions of two of the most specialised ones (setting collapse operators explicitly and varying the number of sub-steps inside each time step of a time-dependent integration). There are also minor restrictions on certain modes of operation, as flagged in the User Manual.

All the necessary data and control parameters are passed to {\tt driveall} through several input files. A computation using this program simply involves (i) specifying the required number of states and other key parameters in the {\tt general\_settings} module; (ii) compiling the program; (iii) preparing or updating the input files as necessary; and (iv) executing the program. The program can be compiled once and for all, as long as the number of states and other key parameters specified in the {\tt general\_settings} module are kept the same.

The {\tt driveall} program reads up to five different input files, of which two must always be provided. The two mandatory files are referred to as the {\tt keyparams} file and the {\tt controlparams} file. Their content is listed in tables~\ref{table:keyparams}, \ref{table:controlparams}, \ref{table:controlparams_123} and \ref{table:controlparams_misc}. For convenience, parameters such as energies and dipole moments, which may be numerous in calculations on large multistate systems, can be specified in an auxiliary {\tt defaultdata} file as an alternative to being listed in the {\tt controlparams} file. The program reads these files using the {\tt namelist} feature of Fortran, as explained in \ref{appendix:example} and \ref{appendix:example2}.
These files must therefore be formatted accordingly; however, no knowledge of Fortran programming is required beyond what is mentioned in this regard in that appendix.

Further examples of the use of {\tt driveall} can be found in the {\tt examples} folder forming part of the distribution. The reader is referred to the User Manual for a full description of all the features of this program. 


\subsection{Running these codes through a bespoke program}
\label{section:bespoke}

\subsubsection{Representation of the density matrix}
\label{section:representation}

As was explained in Section~\ref{section:generalformulation},
the density matrices are stored within these modules as column vectors of $N^2$ real numbers, as per Eq.~(\ref{eq:r_uppertr}): 
a 1D array {\tt rhovec} representing a density matrix is such that {\tt rhovec(1)} contains $\rho_{11}$, {\tt rhovec(2)} contains $\mbox{Re}\,\rho_{12}$, etc.\ (or  $\rho_{00}$, $\mbox{Re}\,\rho_{01}$, etc., if the states are numbered 0, 1, 2,\ldots rather than 1, 2, 3,\ldots, see Section~\ref{section:numbering}). Which components of such vectors correspond to which elements of the density matrix can be found by using
the subroutines {\tt obe\_coher\_index} and {\tt obe\_pop\_index} of the {\tt obe} module.


\subsubsection{The {\tt obefield} and {\tt obecfield} derived types}
\label{section:obefield}

Two Fortran derived variable types, called {\tt obefield} and {\tt obecfield}, are defined in the module {\tt obe}. Variables of this type are used by the {\tt obe} and {\tt mbe} programs for storing and communicating various attributes of the relevant fields, such as their amplitude, wavelength, direction of propagation, detuning, and the Rabi frequencies or dipole moments of the transitions they drive. A full description of these two derived types can be found in the User Manual. No knowledge of these derived types is required for running the codes through the {\tt driveall} program.

\subsubsection{Structure of the program}
\label{section:usingobe}

With the exceptions of the subroutines flagged at the end of this section, using this package will normally involve the steps outlined below.
A Fortran~90 program using the {\tt obe\_steadystate} subroutine for a steady-state calculation is provided in \ref{appendix:example} as an example. A program using the {\tt mbe} module for a propagation calculation is also included in the {\tt examples} directory. Detailed information about using the various user-facing routines provided in this library can be found in the User Manual.

\begin{enumerate}
    \item The driving program should first pass various pieces of information to the {\tt obe} module through a call to the subroutine {\tt obe\_setcsts}, namely
    the frequency offsets of the different states, the rates of spontaneous decay and optionally any additional dephasing rate and any additional collapse operator, as well as the number of fields, whether the weak probe approximation is to be assumed or not, and whether Rabi frequencies or complex electric field amplitudes and dipole moments will be used for defining how each of the fields couples to the different states.
    \item The parameters of each of the fields must then be passed to {\tt obe} through calls to the subroutine {\tt obe\_setfields}. The field identified by the reference number 1 in the corresponding call to {\tt obe\_setfields} is taken to be the probe field if the calculation is to be done within the weak probe approximation.
    \item The root-mean squared velocity of the atoms and the details of the integration over atomic velocities must be passed to {\tt obe} through a call to the subroutine {\tt obe\_set\_Doppler} if a calculation involving a Doppler averaging by numerical quadrature is to be done. A choice of general numerical quadratures is offered by {\tt obe\_set\_Doppler}. Alternatively, the user can upload custom abscissas and weights. 
    \item Unless the applied fields are CW, the details of their temporal envelope must be passed to {\tt mbe} either through a call to {\tt mbe\_set\_envlp} followed by a call to {\tt mbe\_set\_tdfields\_A}, or through a call to {\tt mbe\_set\_tdfields\_B}. The latter makes it possible to use time meshes and define temporal profiles more varied than offered by {\tt mbe\_set\_envlp} and {\tt mbe\_set\_tdfields\_A}.
    \item The relative and absolute accuracy parameters of the DOP853 ODE solver must also be passed to {\tt obe}, through a call to {\tt obe\_set\_tol\_dop853}, if this solver is to be used in the course of the calculation.
    \item Unit numbers for the output of selected elements of the density matrix must be passed to {\tt obe} through a call to {\tt obe\_setoutputfiles} if this option of outputting results is to be used.
    \item The relevant computational routines must then be called for performing the required calculation. 
        The initial populations and (possibly) coherences need to be passed to these various subroutines as input data, with the exceptions mentioned in their detailed descriptions in the User Manual.
    \item A calculation of the density matrix can be followed, if required, by a call to {\tt obe\_susceptibility}, which calculates the complex susceptibility, refractive index and absorption coefficient.
    \item
    For propagation calculations, the program must also include an external subroutine through which {\tt mbe\_propagate\_1} or {\tt mbe\_propagate\_2} can output the results, as described in the User Manual.
\end{enumerate}

The detunings and complex amplitudes of the applied fields initially set by {\tt obe\_setfields} can be reset at a later stage, respectively by calling the subroutines {\tt obe\_reset\_detuning} and {\tt obe\_reset\_cfield} of the {\tt obe} module. This makes it possible, e.g., to calculate refractive indexes and absorption coefficients for a range of detunings or a range of field strengths.

The {\tt obe} module also includes several auxiliary routines which may be of assistance when preparing the input of some of the subprograms mentioned above or processing their output. These are {\tt obe\_coher\_index} and {\tt obe\_pop\_index}, for identifying the relevant elements of a density matrix in the 1D storage mode described in Section~\ref{section:representation}; {\tt obe\_find\_cfield} and {\tt obe\_find\_rabif}, for relating complex electric field amplitudes to complex Rabi frequencies in the definition of Eq.~(\ref{eq:Omegadef}); and {\tt obe\_init\_rho}, for initialising a density matrix in its 1D representation.

None of the initialisation steps listed above are necessary if the only computational routines used would be {\tt obe\_2state}, {\tt obe\_weakprb\_3stladder}, {\tt obe\_weakprb\_4stladder} or {\tt obe\_weakfield}.


\subsection{State numbering}
\label{section:numbering}

The $N$ states included in a calculation are identified by numbers running from 1 to $N$ throughout these code and in the documentation. This numbering is in line with the default indexing of arrays in Fortran. However, the user may choose to use a different numbering system for describing the system in the driving program, e.g., one where the states are identified by a number running from 0 to $N-1$ rather from 1 to $N$. The value of the variable {\tt nmn} set in the {\tt general\_settings} module informs the user-facing {\tt obe} and {\tt mbe} routines of the numbering system used in the external programs calling them: giving a value of $n$ to {\tt nmn} means that the states are numbered $n$, $n+1$, $n+2$,\ldots in the information passed to {\tt obe} and {\tt mbe} by the driving program.
E.g., setting {\tt nmn} to 0 means that the states are numbered 0, 1, 2,\ldots in the arrays passed to these modules by the user, while setting {\tt nmn} to 1 means that the states are instead numbered 1, 2, 3,\ldots Within the {\tt obe}, {\tt mbe} and {\tt ldbl} modules, however, the states are numbered 1, 2 and 3, irrespective of the value of {\tt nmn}. 

For instance, the three states of the systems considered in \ref{appendix:example} could be identified equally well by the numbers 0, 1 and 2, rather than by the numbers 1, 2 and 3. In order to use 0, 1 and 2, the constant {\tt nmn} should be given the value 0 in the {\tt general\_settings} module (and in the {\tt keyparams} file if {\tt driveall} is used). The statements
\begin{lstlisting}
    Gamma_decay(1,2) = 5.0d0
    Gamma_decay(2,3) = 1.0d0
\end{lstlisting}
defining the decay rates in the example given in \ref{appendix:example} should then be replaced by
\begin{lstlisting}
    Gamma_decay(0,1) = 5.0d0
    Gamma_decay(1,2) = 1.0d0
\end{lstlisting}
and similarly for the arrays {\tt Rabif}, {\tt detuning\_fact} and {\tt energ\_f}.


\section{Code reuse}
\label{section:notice}

The {\tt ldbl} and {\tt mbe} modules both contain a copy, in essentially the original form, of the subroutine {\tt DOP853} described in Ref.~\cite{Hairer1993} and published by the University of Geneva \cite{dop853}. 
The {\tt obe} module contains a copy of the subroutine {\tt CLENSHAW\_CURTIS\_COMPUTE} published by J.~Burkardt \cite{Burkardt},
and a Fortran~90 implementation of the {\tt wwerf} function of the CERN Library, which calculates the Faddeeva function \cite{wwerf,Oeftiger2016,ccperrfr}.


\section*{Declaration of competing interest}
The authors have no competing financial interests or personal relationships that could have influenced or may appear to have influenced the work reported in this article.

\section*{Acknowledgements}

The development of these programs has benefited from useful advice and comments from C. S. Adams, I. G Hughes, M. P. A. Jones and T. P. Ogden. The authors also acknowledge the use of the Hamilton HPC Service of Durham University during the course of this work.

\appendix

\section{The Lindblad master equation in the rate equations limit}
\label{appendix:rate}

As mentioned in Section~\ref{section:rate}, a net reduction in the size of the problem can be obtained by propagating only those elements of $\rho$ which belong to a class ${\cal S}$ of elements varying slowly in time. Those that are not propagated form the class ${\cal R}$ of rapidly varying elements. 

Accordingly,
we divide ${\sf r}$ into two column vectors,
${\sf r}_{\cal R}$ and ${\sf r}_{\cal S}$, respectively grouping
the $N_{\cal R}$ elements of ${\sf r}$ belonging to
class ${\cal R}$ and the $N_{\cal S} = N^2-N_{\cal R}$
elements belonging to class ${\cal S}$. Formally,
\begin{align}
{\sf r}_{\calR} & = {\sf R}\,{\sf r}  \\
{\sf r}_{\calS} & = {\sf S}\,{\sf r},
\end{align}
where ${\sf R}$ and ${\sf S}$ are two rectangular matrices, respectively of
size $N_\calR \times N^2$ and $N_\calS \times N^2$. The elements of ${\sf R}$ are
defined by the equation $R_{ij} = \delta_{k(i)j}$,
where $k(i)$ is the index of the $i$-th element of
${\sf r}_\calR$ in the column vector ${\sf r}$. The elements of
${\sf S}$ are defined similarly.
In terms of
the tranposes of these two matrices,
\begin{equation}
{\sf r} = {\sf R}^T{\sf r}_{\calR} + {\sf S}^T{\sf r}_{\calS}.
\end{equation}
We now make the approximation that
the elements of class $\calR$ converge so rapidly to steady values after any
variation of the elements of class $\calS$ that they
can be assumed to remain stationary on the time scale on which
the latter evolve. That is, we set
\begin{equation}
\dot{\sf r}_{\calR} = 0
\label{eq:rR_evol}
\end{equation}
when solving equation (\ref{eq:r_evol}). Equation (\ref{eq:rR_evol}) can also
be written as ${\sf R}\,{\sf L}\,{\sf r} = 0$, from which we deduce that
\begin{equation}
({\sf R}\,{\sf L}\,{\sf R}^T){\sf r}_{\calR} =
 - ({\sf R}\,{\sf L}\,{\sf S}^T){\sf r}_{\calS}.
\label{eq:rR_evol2}
\end{equation}
Given the elements of ${\sf r}_{\calS}$, this equation
determines the elements of
${\sf r}_{\calR}$.
Formally
\begin{equation}
{\sf r}_{\calR} = -({\sf R}\,{\sf L}\,{\sf R}^T)^{-1}({\sf R}\,{\sf L}\,{\sf S}^T)
{\sf r}_{\calS}.
\label{eq:rR_evol3}
\end{equation}
In practice, however, ${\sf r}_{\calR}$ is calculated
by solving equation (\ref{eq:rR_evol2}) as a system of inhomogeneous
linear equations.
We also have
\begin{equation}
\dot{\sf r}_{\calS} = ({\sf S}\, {\sf L}\, {\sf S}^T) {\sf r}_{\calS} +
 ({\sf S}\, {\sf L}\, {\sf R}^T) {\sf r}_{\calR},
\label{eq:rS_evol}
\end{equation}
since $\dot{\sf r}_{\calS} = {\sf S}\,{\sf L}\,{\sf r}$.
Eliminating ${\sf r}_{\calR}$ between equations (\ref{eq:rR_evol3}) and
(\ref{eq:rS_evol}) gives
the equation of motion for the populations and coherences belonging
to class $\calS$:
\begin{equation}
\dot{\sf r}_{\calS} = {\sf L}_{\calS}\,
 {\sf r}_{\calS},
\label{eq:rS_evol2}
\end{equation}
with
\begin{equation}
{\sf L}_{\calS} =
({\sf S}\, {\sf L}\, {\sf S}^T) -
 ({\sf S}\, {\sf L}\, {\sf R}^T) ({\sf R}\,{\sf L}\,{\sf R}^T)^{-1}({\sf R}\,{\sf L}\,{\sf S}^T).
\label{eq:rS_evolmat}
\end{equation}
Contrary to ${\sf L}$, the $N_\calS \times N_\calS$ square matrix ${\sf L}_\calS$
depends on the Rabi frequencies, decoherence rates and and detunings in
a complicated way.
However, it is not difficult to construct this matrix numerically, as
its columns can be obtained one by one by calculating how each unit basis
vector is
transformed by the operator
$({\sf S}\, {\sf L}\, {\sf S}^T) -
 ({\sf S}\, {\sf L}\, {\sf R}^T) ({\sf R}\,{\sf L}\,{\sf R}^T)^{-1}({\sf R}\,{\sf L}\,{\sf S}^T)$.


\section{Calculation of the steady-state density matrix}
\label{appendix:steady}

The linear equations method mentioned in Section~\ref{section:steady} is based on the unit trace property of the density matrix,
\begin{equation}
    \sum_{i=1}^N \rho_{ii} = 1.
\end{equation}
In terms of the elements $r_j$ of a vector ${\sf r}$ of the form of Eq.~(\ref{eq:r_uppertr}),
this property can be formulated as
\begin{equation}
\sum_{j \in {\cal P}} r_j = 1
\label{eq:trace}
\end{equation}
if
the set ${\cal P}$ is defined by the condition that $r_j$ belongs
to ${\cal P}$ if and only if $r_j$ is a population. Let $J$ be one of the elements of this set of indexes, and let ${\cal P}' = {\cal P}\, \backslash \, \{J\}$.
Thus 
\begin{equation}
    r_J = 1 - \sum_{j \in {\cal P}'} r_j.
\label{eq:trace2}
\end{equation}
This relation makes it possible to rearrange the equation ${\sf L}\,{\sf r} = 0$ defining the steady-state density matrix into the equations
\begin{equation}
    \sum_{j \in {\cal P}'} (L_{ij} - L_{iJ})\,r_j +
    \sum_{j \not\in {\cal P}} L_{ij}\,r_j =
    -L_{iJ}, \qquad i = 1,\ldots,N^2.
\label{eq:newsystem}
\end{equation}
Moreover, Eq.~(\ref{eq:trace2}) also makes the equation
\begin{equation}
    \dot{r}_J = \sum_{j} L_{Jj}\, r_j = 0
\end{equation}
redundant with the rest of the original system since $\dot{r}_J$ is necessarily zero if $\dot{r}_j$ is zero for all the $j$'s belonging to ${\cal P}'$. The equation for $i=J$ can thus be removed from Eq.~(\ref{eq:newsystem}). The other equations then form an inhomogeneous system of $N^2-1$ linear equations in the $N^2-1$ unknowns $r_j$ ($j\not= J$), as expressed by Eq.~(\ref{eq:steadylinear}).



This system of equations can be solved numerically by standard methods. However, it can also be solved by the following method, which is well suited to calculations of the steady-state density matrix with Doppler broadening, as we now explain.

The structure of the optical Bloch equations ensures that the $L_{iJ}$'s forming the right-hand sides of Eq.~(\ref{eq:newsystem}) do not depend on detunings, and that the other $L_{ij}$'s depend at most linearly on them. The column vector ${\sf b}$ is thus constant in the atomic velocity $v$, while the matrix ${\sf L}'$ varies linearly with $v$. We set, accordingly,
\begin{equation}
{\sf L}' = {\sf L}'_0 + v\, {\sf L}'_1,
\label{eq:Lprimev}
\end{equation}
where ${\sf L}'_0$ and ${\sf L}'_1$ do not depend on $v$.
The two matrices ${\sf L}'_0$ and ${\sf L}'_1$ define the generalized eigenvalue problem
\begin{equation}
    {\sf L}'_0 {\sf x} = \mu\, {\sf L}'_1{\sf x},
\end{equation}
where $\mu$ is a (normally complex) generalized eigenvalue.
Since ${\sf L}'_0$ and ${\sf L}'_1$ are $({\cal N}-1)\times ({\cal N}-1)$ matrices,
the span of the solution vectors ${\sf x}$ is a space of dimension ${\cal M}\leq {\cal N}-1$. It is thus possible to find ${\cal M}$ eigenvectors ${\sf x}_1$, ${\sf x}_2$,\ldots, ${\sf x}_{\cal M}$ forming a basis for this space. With $\mu_j$ denoting the corresponding eigenvalues,
\begin{equation}
    {\sf L}'_0 {\sf x}_j = \mu_j\, {\sf L}'_1{\sf x}_j,  \quad
    j = 1,\ldots,{\cal M}.
    \label{eq:righeigen}
\end{equation}
To each eigenvector ${\sf x}_j$ can be associated a left eigenvector ${\sf y}_j$ such that
\begin{equation}
    {\sf y}_j^\dagger{\sf L}'_0 = \mu_j\, {\sf y}_j^\dagger{\sf L}'_1, \quad
    j = 1,\ldots,{\cal M}
    \label{eq:lefteigen}
\end{equation}
and
\begin{equation}
     {\sf y}_i^\dagger{\sf L}'_1 {\sf x}_j = \delta_{ij}.
     \label{eq:biorthogonality}
\end{equation}
The solution ${\sf r}'$ of Eq.~(\ref{eq:steadylinear}) can be written as the sum of a linear combination of the eigenvectors ${\sf x}_j$'s and of a vector ${\sf r}'_0$ biorthogonal to all the left eigenvectors: 
\begin{equation}
{\sf r}' = \sum_j c_j {\sf x}_j + {\sf r}'_0,
\end{equation}
with ${\sf r}'_0$ being such that
\begin{equation}
    {\sf y}_j^\dagger{\sf L}'_1{\sf r}'_0 = 0, \quad
    j = 1,\ldots,{\cal M}.
    \label{eq:null}
\end{equation}
(Formally, $c_j = {\sf y}_j^\dagger{\sf L}'_1{\sf r}'$ and ${\sf r}'_0 = {\sf r}' - \sum_j c_j {\sf x}_j$.)
Combining the above equations yields
\begin{equation}
    c_j = \frac{{\sf y}_j^\dagger{\sf b}}{v+\mu_j}, \quad j = 1,\ldots,{\cal M}
\end{equation}
and
\begin{equation}
    {\sf L}'_0{\sf r}'_0 = {\sf b}-\sum_j (v+\mu_j)c_j{\sf L}'_1{\sf x}_j.
    \label{eq:laststep}
\end{equation}
Solving Eqs.~(\ref{eq:righeigen}) and (\ref{eq:lefteigen}) for the eigenvalues $\mu_j$ and the corresponding right and left eigenvectors is a standard numerical problem, as is solving Eq.~(\ref{eq:laststep}) for the vector ${\sf r}'_0$. (In the present programs, this calculation is done by first reverting to a formulation of the density matrix in terms of real populations and complex coherences, and working with the complex matrices and complex vectors corresponding to ${\sf L}'$, ${\sf r}'$ and ${\sf b}$ in that formulation.) Altogether, the calculation yields each of the elements of ${\sf r}_{\rm st}$ as a sum of partial fractions with constant numerators and denominators linear in $v$:
\begin{equation}
    ({\sf r}_{\rm st})_i = \sum_j \frac{\alpha_{ij}}{v+\mu_j},
\end{equation}
where the $\alpha_{ij}$'s are constants. As is well known, expressions of this form are readily amenable to an analytical averaging over a Maxwellian distribution of velocities \cite{Gea-Banacloche1995}, and indeed, expanding coherences in partial fractions of this form is a standard approach in few-state calculations based on the weak probe approximation. The method outlined in this appendix generalises this approach to multi-state, multi-fields systems treated beyond the weak probe approximation.

Doppler averaging is based on the following identities, where
$ \eta_{j} = -\mu_j/u$ and, as defined in Section~\ref{section:Doppler}, $w(\cdot)$ is the Faddeeva function:
\begin{align}
   \int_{-\infty}^\infty \, \frac{f_{\rm M}(v) \, {\rm d}v}{v+\mu_j} &= 
    \frac{1}{u \sqrt{\pi}}
    \int_{-\infty}^\infty \, \frac{\exp(-\eta^2) \, {\rm d}\eta}{\eta-\eta_{j}}
    \label{eq:F1} \\ &=
     \frac{1}{u \sqrt{\pi}}
          \begin{cases}
i\pi w(\eta_{j}) & \mbox{if Im~$\eta_{j} > 0$}, \\
[i\pi w(\eta^*_{j})]^* & \mbox{if Im~$\eta_{j} < 0$}.
\end{cases}
\label{eq:F3}
\end{align}
The case $\mbox{Im}~\eta_j = 0$ does not need to be considered as the eigenvalues $\mu_j$ always have a non-zero imaginary part for any pair of matrices ${\sf L}'_0$ and ${\sf L}'_1$ arising from the optical Bloch equations.

Organising the calculations along similar lines may also lead to a significant speed up in computations of the steady state density matrix for multiple values of one of the detunings. For such calculations, Eq.~(\ref{eq:Lprimev}) would be replaced by the equation
\begin{equation}
    {\sf L}' = {\sf {\tilde L}}'_0 + \Delta_\alpha {\sf {\tilde L}}'_1,
\end{equation}
where the matrices ${\sf {\tilde L}}'_0$ and ${\sf {\tilde L}}'_1$ do not depend on $\Delta_\alpha$. Following the above procedure then results in a density matrix of the form
\begin{equation}
    ({\sf r}_{\rm st})_i(\omega) = \sum_j \frac{{\tilde \alpha}_{ij}}{\Delta_\alpha+{\tilde \mu}_j},
    \label{eq:expansion2}
\end{equation}
where ${\tilde \alpha}_{ij}$ and ${\tilde \mu}_j$ are constants.
The only potentially CPU intensive step in this approach is the calculation of the generalized eigenvalues and eigenvectors of the matrix pair $({\sf {\tilde L}}'_0,{\sf {\tilde L}}'_1)$, which does not need to be repeated for each value of $\Delta_\alpha$.


\section{The optical Bloch equations in the weak probe approximation}
\label{appendix:weak}

For simplicity, we will only consider the case where the amplitude of the probe field, ${\cal E}_{\rm p}$, is real. The final results --- Eqs.~(\ref{eq:r_evol0}) and (\ref{eq:r_evol1}) --- are easily generalised to the case of a complex amplitude, and the program is organised in such a way that the weak probe approximation is correctly implemented whether the amplitude of the probe field is real or complex.

We assume that the coherences are all initially
zero. The populations then vary with ${\cal E}_{\rm p}$ only through
terms quadratic or of higher order in ${\cal E}_{\rm p}$. The populations will therefore
vary little if the probe field is very weak.
The essence of the weak probe approximation is to integrate Eq.~(\ref{eq:r_evol}) only to the leading (non vanishing) order in ${\cal E}_{\rm p}$. This is done to first order in ${\cal E}_{\rm p}$ within the {\tt obe} module.

Implementing this approximation first requires a consideration of Eq.~(\ref{eq:r_evol}) in the limit of a vanishing probe field (${\cal E}_{\rm p} \rightarrow 0$).
The elements of the density matrix divide into two classes in
that limit, 
namely the populations and the coherences which take on non-zero values either initially or at later
times (class ${\cal A}$),
and the coherences which are initially zero and remain zero at all times (class ${\cal B}$).
(The elements of class ${\cal A}$ may vary in time even when ${\cal E}_{\rm p} = 0$, e.g., because of spontaneous decay or because of an interaction with a field other than the probe field.)
We can thus form
the column vector ${\sf r}$ by
concatenating the column vectors formed by the respective
populations and coherences, ${\sf r}_{\cal A}$ and ${\sf r}_{\cal B}$:
\begin{equation}
{\sf r} \equiv
\begin{pmatrix}
{\sf r}_{\cal A} \\ {\sf r}_{\cal B}
\end{pmatrix}.
\end{equation}
Accordingly, Eq.~(\ref{eq:r_evol}) takes on the form
\begin{equation}
\begin{pmatrix}
\dot{\sf r}_{\cal A} \\ \dot{\sf r}_{\cal B}
\end{pmatrix} =
\begin{pmatrix}
{\sf L}^{}_{\cal AA} & {\sf L}^{}_{\cal AB} \\
{\sf L}^{}_{\cal BA} & {\sf L}^{}_{\cal BB}
\end{pmatrix}
\begin{pmatrix}
{\sf r}_{\cal A} \\ {\sf r}_{\cal B}
\end{pmatrix},
\label{eq:block1}
\end{equation}
where the blocks ${\sf L}^{}_{\cal AA}$, ${\sf L}^{}_{\cal AB}$, ${\sf L}^{}_{\cal BA}$
and ${\sf L}^{}_{\cal BB}$ are square or rectangular matrices. Since the optical Bloch equations are linear in the Rabi frequencies,
each of these blocks is constant or linear in ${\cal E}_{\rm p}$:
\begin{equation}
{\sf L}_{\cal AA} = {\sf L}^{(0)}_{\cal AA} + {\cal E}_{\rm p}\, {\sf L}^{(1)}_{\cal AA},
\label{eq:blocks}
\end{equation}
where the matrices
${\sf L}^{(0)}_{\cal AA}$ and ${\sf L}^{(1)}_{\cal AA}$ are constant in
${\cal E}_{\rm p}$, and similarly for the other blocks. Due to this dependence in the probe field, both ${\sf r}_{\cal A}$ and ${\sf r}_{\cal B}$ may depend in a complicated way on ${\cal E}_{\rm p}$. 

In general, a perturbative expansion of these two vectors reads
\begin{align}
{\sf r}_{\cal A} & =  {\sf r}^{(0)}_{\cal A} + {\cal E}_{\rm p}\, {\sf r}^{(1)}_{\cal A} +
{\cal E}_{\rm p}^2\, {\sf r}^{(2)}_{\cal A} + \cdots \\
{\sf r}_{\cal B} & =  {\sf r}^{(0)}_{\cal B} + {\cal E}_{\rm p}\, {\sf r}^{(1)}_{\cal B} +
{\cal E}_{\rm p}^2\, {\sf r}^{(2)}_{\cal B} + \cdots
\end{align}
where the vectors ${\sf r}^{(k)}_{\cal A}$'s and
${\sf r}^{(k)}_{\cal B}$'s do not depend on ${\cal E}_{\rm p}$.
We can immediately see that ${\sf r}^{(0)}_{\cal B} = 0$, since, by construction, the vector ${\sf r}_{\cal B}$ groups all the elements of ${\sf r}$
which are zero at all times in the limit ${\cal E}_{\rm p} \rightarrow 0$. Moreover, the elements of ${\sf r}_{\cal A}$ are non-zero even in this limit, whether they are initially non-zero or whether they acquire a non-zero value as $t$ increases. The leading term in the perturbative expansion of ${\sf r}_{\cal A}$ is thus the term of order 0 in ${\cal E}_{\rm p}$. Also, ${\sf L}^{(0)}_{\cal BA}$ must be zero, as otherwise ${\sf r}_{\cal B}$ would not be identically zero in the ${\cal E}_{\rm p} \rightarrow 0$ limit. 
Retaining the terms of lowest order in ${\cal E}_{\rm p}$ thus implies setting ${\sf r}^{(k)}_{\cal A} = 0$ for $k \not= 0$ and ${\sf r}^{(k)}_{\cal B} = 0$ for $k \not= 1$, and finding these vectors as solutions of the equations
\begin{align}
\label{eq:r_evol0}
\dot{\sf r}_{\cal A}^{(0)} &= {\sf L}_{\cal AA}^{(0)}\,{\sf r}_{\cal A}^{(0)}\\
\label{eq:r_evol1}
\dot{\sf r}_{\cal B}^{(1)} &= {\sf L}_{\cal BA}^{(1)}\,{\sf r}_{\cal A}^{(0)} +
{\sf L}_{\cal BB}^{(0)}\,{\sf r}_{\cal B}^{(1)}.
\end{align}
(The replacement of the diagonal block ${\sf L}^{}_{\cal BB}$ by its
 zero-${\cal E}_{\rm p}$ limit, ${\sf L}^{(0)}_{\cal BB}$, ensures
that ${\sf r}_{\cal B}^{(1)}$ remains linear in ${\cal E}_{\rm p}$.)


In summary, the weak probe approximation amounts to integrating the equations (\ref{eq:r_evol0}) and (\ref{eq:r_evol1}) rather than Eq.~(\ref{eq:r_evol}). I.e., it amounts to replacing the matrix ${\sf L}$ by the matrix
$$
\begin{pmatrix}
{\sf L}_{\cal AA}^{(0)} & 0 \\
{\sf L}_{\cal BA}^{(1)} & 
{\sf L}_{\cal BB}^{(0)}
\end{pmatrix}.
$$
The key steps in implementing this approximation is to construct the matrix ${\sf L}_{\cal AA}^{(0)}$ and allocate the elements of ${\sf r}$ to either ${\sf r}_{\cal A}$ or
${\sf r}_{\cal B}$. Within {\tt obe}, this is done by an iterative search for the  elements of ${\sf r}$ directly or indirectly coupled by ${\sf L}$ to the initial non-zero populations when
${\cal E}_{\rm p} = 0$.

\section{Weak probe calculations for a single field}
\label{appendix:single}

This appendix addresses the case of a single, weak CW field. We will assume that this field dipole-couples one set of states, ${\cal G}_1$, to another set of states, ${\cal G}_2$, higher in energy. Each of the latter decays spontaneously at a state-dependent rate $\Gamma_j$. The former are stable. We describe the field by way of Eqs.~(\ref{eq:Efirst}) and (\ref{eq:Esecond}), although without specifying the subscript $\alpha$ for economy of notation (it is understood that $\alpha = 1$). We make the weak probe approximation and assume that only the states belonging to group ${\cal G}_1$ are populated. Thus $\rho_{ii} = 0$ if $i \in {\cal G}_2$ and
\begin{equation}
    \sum_{j \in {\cal G}_1} \rho_{jj} = 1.
\end{equation}
Eq.~(\ref{eq:Lindblad}) simplifies considerably in that limit: $\rho_{ij} \equiv 0$ if $i$ and $j$ both belong to ${\cal G}_1$ or both belong to ${\cal G}_2$, whereas
\begin{align}
    \dot{\rho}_{ij} = i \left[
\left(
\delta \omega^{(j)} - \delta\omega^{(i)} + \Delta
\right)\rho_{ij} + \frac{\Omega_{ij}}{2}\,\rho_{jj}
\right] - \frac{\Gamma_i}{2}\, \rho_{ij} -
\gamma^{}_{ij} \rho_{ij}
\end{align}
when $i \in {\cal G}_2$ and $j \in {\cal G}_1$. The decoherence rates $\gamma^{}_{ij}$ account for dephasing mechanisms not contributing to the decay rates $\Gamma_i$, such as random phase jumps of the field contributing to its frequency width \cite{Wodkiewicz1979} and collisional broadening. Typically,
\begin{equation}
\gamma_{ij} = \gamma_{ij}^{\rm coll} + 2\pi \Delta\nu,
\end{equation}
where $\gamma_{ij}^{\rm coll}$ is the decay rate of the coherence $\rho_{ij}$ due to collisions and $\Delta\nu$ is the frequency width of the field (full width at half maximum).

We refer the energies of the states to either an energy $\hbar\omega_{\rm ref}(1)$ or $\hbar \omega_{\rm ref}(2)$ depending on whether they belong to group ${\cal G}_1$ or group ${\cal G}_2$. Thus
\begin{align}
    \delta\omega^{(j)} &= \omega^{(j)} -\omega_{\rm ref}(1) \quad \mbox{if $j \in {\cal G}_1$}, \\
    \delta\omega^{(i)} &= \omega^{(i)} -\omega_{\rm ref}(2) \quad \mbox{if $i \in {\cal G}_2$}.
\end{align}
Moreover
\begin{equation}
    \Delta = \omega - [\omega_{\rm ref}(2) - \omega_{\rm ref}(1)],
\end{equation}
and therefore, in the above equation,
\begin{equation}
    \delta \omega^{(j)} - \delta\omega^{(i)} + \Delta \equiv \omega - \left[\omega^{(i)} - \omega^{(j)}\right].
\end{equation}
Moreover, since we defined ${\cal G}_2$ as containing states higher in energy than the states belonging to ${\cal G}_1$, 
\begin{equation}
\Omega_{ij} = {\cal E}\,
 \langle\, i\, |\,
\hat{\text{\boldmath{$\epsilon$}}} \cdot \hat{\bf D} \,|\, j\, \rangle/\hbar.
\label{eq:weakOmega}
\end{equation}

Setting $\dot{\rho}_{ij} = 0$ yields the steady state coherences:
\begin{equation}
    \rho_{ij} = \frac{i}{2}\,\frac{ \Omega_{ij}\rho_{jj}}{\gamma^{\rm tot}_{ij} - i\, \Delta_{ij}}
    \label{eq:weakrhoij}
\end{equation}
with
\begin{equation}
    \gamma^{\rm tot}_{ij} = \Gamma_i/2 + \gamma^{}_{ij}, \qquad
    \Delta_{ij} = \omega - \left[\omega^{(i)} - \omega^{(j)}\right].
    \label{eq:Definegamma}
\end{equation}
Given Eqs.~(\ref{eq:weakOmega}) and (\ref{eq:weakrhoij}), Eq.~(\ref{eq:chi}) yields a particularly simple result for
the corresponding complex susceptibility:
\begin{equation}
    \chi(\omega_1) = \frac{i N_{\rm d}}{\hbar\epsilon_0}\, \sum_{i\in {\cal G}_2} \sum_{j \in {\cal G}_1} \, \frac{
    |\langle\, i\, |\,
\hat{\text{\boldmath{$\epsilon$}}}_\alpha \cdot \hat{\bf D} \,|\, j\, \rangle|^2}
{\gamma^{\rm tot}_{ij} - i \Delta_{ij}} \,\rho_{jj}.
\end{equation}
Note that the full width at half maximum of the resonance peak at $\Delta_{ij} = 0$ is twice the total dephasing rate $\gamma^{\rm tot}_{ij}$. E.g., to obtain a collisional width of $\Gamma^{\rm coll}_{ij}$ (full width at half maximum in angular frequency), the dephasing rate $\gamma_{ij}^{\rm coll}$ must be set equal to $\Gamma^{\rm coll}_{ij}/2$.

Doppler averaging $\chi(\omega_1)$ then amounts to a simple application of Eqs.~(\ref{eq:F1}) and (\ref{eq:F3}), since
\begin{align}
    \frac{1}{u\sqrt{\pi}}\,\int_{-\infty}^\infty \, \frac{\exp(-v^2/u^2) \, {\rm d}v}{\gamma_{ij}^{\rm tot} - i \Delta_{ij} + i k v} = 
    \frac{1}{iu k \sqrt{\pi}}
    \int_{-\infty}^\infty \, \frac{\exp(-\eta^2) \, {\rm d}\eta}{\eta-\eta_{ij}}
\end{align}
with $\eta_{ij} = \left(\Delta_{ij} + i\, \gamma_{ij}^{\rm tot}\right)/(uk)$.
Therefore
\begin{align}
    \frac{1}{u\sqrt{\pi}}\,\int_{-\infty}^\infty \, \frac{\exp(-v^2/u^2) \, {\rm d}v}{\gamma_{ij}^{\rm tot} - i \Delta_{ij} + i k v} = 
    \sqrt{\pi}\,w(\eta_{ij})/(uk).\nonumber \\
\end{align}


\section{Example of steady state calculation}
\label{appendix:example}

This appendix illustrates how the {\tt obe} codes can be used for calculating the steady state density matrix for a ladder system of three states, states~1, 2 and 3, with $\hbar\omega^{(1)} < \hbar\omega^{(2)} < \hbar\omega^{(3)}$. States~1 and 2 are coupled to each other by field~1  (the ``probe field") and states~2 and 3 by field~2 (the ``coupling field"). Within the rotating wave approximation, the Hamiltonian of this systems is represented by the following matrix in the $\{|1\rangle,|2\rangle,|3\rangle\}$ basis, 
\begin{equation}
    {\sf H}' = \hbar \begin{pmatrix}
    \delta\omega^{(1)} & -{\Omega}_{12}/2 & 0 \\
    -{\Omega}_{12}^*/2 & \delta\omega^{(2)} -\Delta_1 & -{\Omega}_{23}/2 \\
    0 &  -{\Omega}_{23}^*/2 & \delta\omega^{(3)} -\Delta_1-\Delta_2
    \end{pmatrix},
\label{eq:Hprimeladder}
\end{equation}
where $\Delta_1$ and $\Delta_2$ are given by Eqs.~(\ref{eq:delta1}) and (\ref{eq:delta2}).
It is assumed that state~3 decays to state~2 and state~2 to state~1, the corresponding decay rates being $\Gamma_{32}$ and $\Gamma_{21}$, respectively, and that inhomogeneous broadening can be neglected. Specifically, we take $\delta\omega^{(1)} = \delta\omega^{(2)} = \delta\omega^{(3)} = 0$, $\Delta_1 = 2\pi \times 5$~MHz, $\Delta_2 =0$, $\Omega_{12} = 2\pi \times 5$~MHz, $\Omega_{23} = 2\pi \times 10$~MHz, $\Gamma_{12} = 2\pi\times 5$~MHz and $\Gamma_{23} = 2\pi\times 1$~MHz, and we calculate the steady state with the subroutine {\tt obe\_steadystate}.
As there are three states in the problem, the {\tt general\_settings} module must give a value of 3 to the variable ${\tt nst}$.

We first show how the steady state density matrix could be obtained by running the code through the {\tt driveall} program. We then give an example of a bespoke program doing the same calculation. Copies of these files are provided in the {\tt examples} folder.

\subsection*{1. Using the {\tt driveall} program}

Calculating this density matrix using the {\tt driveall} program requires two input files, namely the {\tt keyparams} and the {\tt controlparams} files, formatted as illustrated by the examples below.  
Apart from possible comments and blank lines, each of these two files must start with an ampersand symbol followed by the name of the respective {\tt namelist} structure ({\tt keyparams} for the {\tt keyparams} file, {\tt controlparams} for the {\tt controlparams} file) and must end with a slash. Each input value must be provided in the form of a Fortran assignment statement (e.g., {\tt i = 1} if {\tt i} is an {\tt integer} variable, {\tt v = 1.0d0} if {\tt v} is a {\tt double precision} variable, {\tt s = 'something'} if {\tt s} is a {\tt character} variable). Input values can be provided in any order and do not need to be all present. Strings of characters starting with an exclamation mark are taken to be comments and are ignored, as are blank lines.

The following file could be used as the {\tt keyparams} file for that calculation. This file is read by {\tt driveall} from the standard input stream. It specifies several key parameters and the name of the {\tt controlparams} file.

\lstset{
basicstyle=\footnotesize\ttfamily,
columns=flexible,
breaklines=true,
breakatwhitespace=true
}

\begin{footnotesize}
\begin{verbatim}
&keyparams
   nstates = 3   ! Number of states.
   nmin = 1      ! I.e., the states are numbered 1, 2, 3.
   nfields = 2   ! Number of fields.
   icmplxfld = 0 ! Indicates that the field amplitudes 
      ! and Rabi frequencies will be specified as real
      ! numbers, not as complex numbers.
                  
!  Name of the controlparams files:
   filename_controlparams = 'example_c.dat'
/
\end{verbatim}
\end{footnotesize}
The corresponding {\tt controlparams} file could be taken to be as follows.\footnote{
Since the three states are assumed to coincide in energy with their respective reference energy level, the frequency offset of each of the states is zero here, which is the default value of these quantities within the {\tt driveall} program. Although this is not necessary, these frequency offsets are specified to be zero in this example, for clarity.}
\lstset{
basicstyle=\footnotesize\ttfamily,
columns=flexible,
breaklines=true,
breakatwhitespace=true
}

\begin{footnotesize}
\begin{verbatim}
&controlparams
   icalc = 2    ! Tells driveall to calculate the
                ! steady state.
   iRabi = 1    ! The Rabi frequencies will be specified.
   inoncw = 0   ! The fields are CW.
   iweakprb = 0 ! The weak probe approximation is not made.
   iDoppler = 0 ! No Doppler averaging.

!  Rabi frequencies, in units of (2 pi) x MHz:
   Rabif(2,1,1) =  5.0d0
   Rabif(3,2,2) = 10.0d0

!  Decay rates, in units of (2 pi) x MHz:
   Gamma_decay_f(1,2) = 5.0d0
   Gamma_decay_f(2,3) = 1.0d0

!  Frequency offset of each of the states, in units of
!  (2 pi) x MHz:
   energ_f(1) = 0.0d0
   energ_f(2) = 0.0d0
   energ_f(3) = 0.0d0

!  The factors multiplying the detunings in the Hamiltonian
!  (only the non-zero values need to be specified):
   detuning_fact(2,1) = -1.0d0
   detuning_fact(3,1) = -1.0d0
   detuning_fact(3,2) = -1.0d0

!  The detunings, in units of (2pi) x MHz:
   detuning(1) = 5.0d0
   detuning(2) = 0.0d0
/
\end{verbatim}
\end{footnotesize}

\noindent Running {\tt driveall} with these input files produces the following output:
\lstset{
basicstyle=\footnotesize\ttfamily,
columns=flexible,
breaklines=true,
breakatwhitespace=true
}
\begin{footnotesize}
\begin{verbatim}
   i   j   Re rho(i,j)   Im rho(i,j)

   1   1   5.85372E-01   0.00000E+00
   1   2  -3.36553E-02  -1.98712E-01
   2   2   1.98712E-01   0.00000E+00
   1   3  -6.03183E-02   1.81884E-01
   2   3  -1.51570E-01  -2.15916E-02
   3   3   2.15916E-01   0.00000E+00
\end{verbatim}
\end{footnotesize}

\subsection*{2. Using a bespoke program}

The program listed below calculates and writes out the steady state value of $\rho_{12}$ for the same parameters. Running it produces the following output:
\lstset{
basicstyle=\footnotesize\ttfamily,
columns=flexible,
breaklines=true,
breakatwhitespace=true
}
\begin{footnotesize}
\begin{verbatim}
 rho(1,2) = -3.36553E-02  -1.98712E-01
\end{verbatim}
\end{footnotesize}


\lstset{
basicstyle=\footnotesize\ttfamily,
columns=flexible,
breaklines=true,
breakatwhitespace=true
}

\begin{footnotesize}
\begin{verbatim}
      program example
 
!  Modules directly used by this program:
      use general_settings
      use obe
 
!  Declare all the variables. The type obecfield is
!  defined in the obe module. The variable nst (the number
!  of states) is defined in the general_settings module.
      implicit none
      type(obecfield) :: coupling_field, probe_field
      double precision, dimension(nst,nst) :: Gamma_decay_f
      double precision, dimension(nst*nst) :: rhovec
      double precision, dimension(nst) :: energ_f
      integer :: ioption, iRabi, iweakprb, mim, mre, nfields
 
!  Properties of the probe field
      probe_field%detuning = 5.0d0
      probe_field%detuning_fact(1) =  0.0d0
      probe_field%detuning_fact(2) = -1.0d0
      probe_field%detuning_fact(3) = -1.0d0
      probe_field%Rabif = (0.0d0 , 0.0d0)
      probe_field%Rabif(1,2) = (5.0d0 , 0.0d0) 
 
!  Properties of the coupling field
      coupling_field%detuning = 0.0d0
      coupling_field%detuning_fact(1) =  0.0d0
      coupling_field%detuning_fact(2) =  0.0d0
      coupling_field%detuning_fact(3) = -1.0d0
      coupling_field%Rabif = (0.0d0 , 0.0d0)
      coupling_field%Rabif(2,3) = (10.0d0 , 0.0d0)
 
!  Frequency offset of each of the states. Here they are
!  zero since the three states are assumed to coincide
!  in energy with their respective reference
!  energy level.
      energ_f(1) = 0.0d0
      energ_f(2) = 0.0d0
      energ_f(3) = 0.0d0
 
!  State 2 decays to state 1 and state 3 decays to state 2.
!  Corresponding decay rates:
      Gamma_decay_f = 0.0d0
      Gamma_decay_f(1,2) = 5.0d0 
      Gamma_decay_f(2,3) = 1.0d0
 
!  Initialisation: Key parameters are first passed to obe
!  through a call to obe_setcsts. The properties of field 1
!  and of field 2 are then passed through calls
!  to obe_setfields.
      nfields = 2    ! Number of fields
      iweakprb = 0   ! 0 means that the weak probe 
                     ! approximation is not made
      iRabi = 1      ! 1 means that the Rabi frequencies are
                     ! provided directly rather than through
                     ! dipole matrix elements and
                     ! field amplitudes.
      call obe_setcsts(energ_f,Gamma_decay_f,nfields,    &
                         iweakprb,iRabi)
      call obe_setfields(1,probe_field) 
      call obe_setfields(2,coupling_field)
 
!  Calculation of the steady state density matrix. The
!  result is returned by obe_steadystate through the
!  1D array rhovec.
      ioption = 1    ! See the description of
                     ! obe_steadystate for that option.
      call obe_steadystate(rhovec,ioption)
 
!  Find out which components of rhovec correspond to the
!  real and imaginary parts of rho_12, and print this
!  coherence.
      call obe_coher_index(1,2,mre,mim)
      print 1000,rhovec(mre),rhovec(mim)
 1000 format(1x,'rho(1,2) = ',2(1pe12.5,2x))
 
      end program example
\end{verbatim}
\end{footnotesize}


\section{Example of propagation calculation}
\label{appendix:example2}

\begin{figure}[tb]
	\centering
	\includegraphics[width=8.5cm]{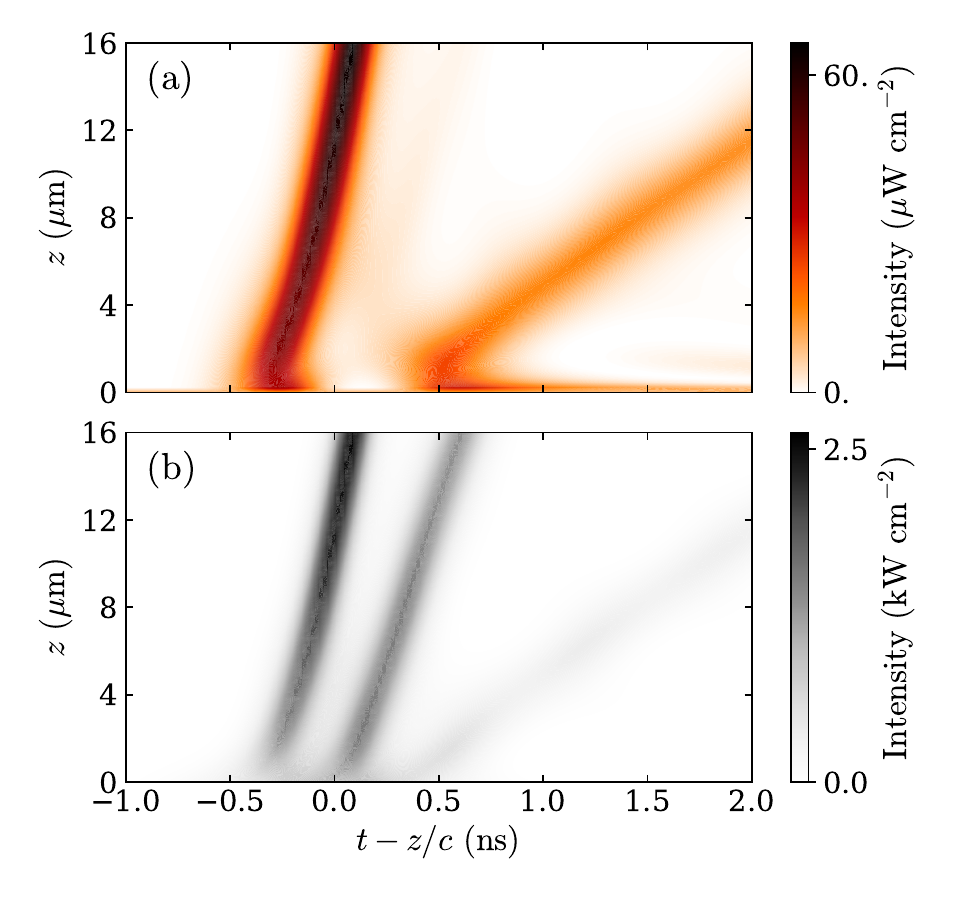}
 \caption{Nascent quasisimultons in a 3-state model, calculated as described in \ref{appendix:example2}. (a): Probe field. (b): Coupling field.}
\label{fig:propagation}
\end{figure}

This second example illustrates the use of the {\tt mbe} module in calculations of light propagation in a nonlinear medium. The calculation reproduces some of the 3-state results reported in Ref.~\cite{Ogden2019}. The medium is a dense ${}^{85}$Rb vapour. The three states in question are the 5$\,^2\mbox{S}_{1/2}$ ground state, the 5$\,^2\mbox{P}_{1/2}$ state and the 5$\,^2\mbox{P}_{3/2}$ state. The D1 transition is addressed by a weak CW field (the probe field), the D2 transition by a strong Gaussian pulse propagating in the same direction as the CW field (the coupling field). The peak intensity of this pulse is so high, 1~kW~cm$^{-2}$, that it splits into three soliton-like pulses travelling over large distances through the medium \cite{McCall1967}. The intensity of the applied CW field is only 10~$\mu$W~cm$^{-2}$, which is normally far too low for the formation of solitons: instead, this field would normally be absorbed rapidly by the vapour. However, a nonlinear interaction between the two fields mediated by the atoms gives rise to three pulses at the D1 wavelength which co-propagate with the strong soliton-like pulses formed at the D2 wavelength, forming as many quasi-simultons \cite{Ogden2019}. This nonlinear effect is illustrated by Fig.~\ref{fig:propagation}, which shows the results generated by the {\tt driveall} program when run with the input data given below.

Both fields are assumed to be $\pi$-polarized in the $z$-direction. The relevant dipole moments are $\langle 5\,^2\mbox{S}_{1/2} \,||\, er\, ||\, 5\,^2\mbox{P}_{1/2} \, \rangle/\sqrt{3}$ for the D1 transition and $\langle 5\,^2\mbox{S}_{1/2} \,||\, er\, ||\, 5\,^2\mbox{P}_{3/2} \, \rangle/\sqrt{3}$ for the D2 transition, where $\langle 5\,^2\mbox{S}_{1/2}\, ||\, er\, ||\, 5\,^2\mbox{P}_{1/2} \,\rangle = 2.537\times 10^{-29}$~C~m \cite{Zentile2015,Volz1996} and 
 $\langle 5\,^2\mbox{S}_{1/2}\, ||\, er\, ||\, 5\,^2\mbox{P}_{3/2} \,\rangle = 3.58425\times 10^{-29}$~C~m \cite{Steck}. 
The spontaneous decay rates of the $5\,^2\mbox{P}_{1/2}$ and $5\,^2\mbox{P}_{3/2}$ states are, respectively, $2\pi \times 5.746$~MHz and $2\pi \times 6.0666$~MHz for the $5\,^2\mbox{P}_{3/2}$ state \cite{Zentile2015,Volz1996}. The atomic number density, $1.96\times 10^{21}$~m$^{-3}$, corresponds to a vapour temperature of 220~$^{\circ}$C; however, Doppler broadening is neglected in this example so as to avoid unnecessarily long execution times. The Maxwell-Bloch equations are integrated in time using Butcher's 5th order Runge-Kutta method and in space using a predictor-corrector method combining the third order Adams-Bashford method and the fourth order Adams-Moulton method, initiated by calculation with smaller spatial steps using the classic fourth order Runge-Kutta rule. The applied fields are read from a file called {\tt appliedfields.dat}. This file, as well as the {\tt controlparams} file listed below and the corresponding {\tt keyparams} file can be found in the {\tt examples} directory included in this distribution. Running the {\tt driveall} program with these input data produces a file called {\tt outamplitudes.dat} containing the complex amplitudes of the propagated fields as functions of position and time (however, see \ref{appendix:podman} for adapting the program listed below to being run through Podman). These complex amplitudes are transformed into the corresponding intensities by the program used for plotting Fig.~\ref{fig:propagation}, which is also included in the {\tt examples} directory.

\lstset{
basicstyle=\normalsize\ttfamily,
columns=flexible,
breaklines=true,
breakatwhitespace=true
}

\begin{footnotesize}
\begin{verbatim}
&controlparams
   icalc = 3  ! Tells driveall to integrate the
              ! Maxwell-Bloch equations.
   iRabi = 0  ! Dipoles moment rather than Rabi frequencies
              ! are specified.
   inoncw = 1 ! Non-CW fields.
   
!  Use obe_set_tdfields_B to define the time-dependent
!  amplitudes of the applied fields, and read these
!  amplitudes from the file appliedfields.dat:
   itdfieldsAorB = 2
   filename_tdamps_in = 'appliedfields.dat'
   n_time_steps = 100  ! Number of time steps.

!  Choice of integration rule for the time integration and
!  number of intermediate steps between each mesh point:
   imethod = 5  ! Butcher's 5-th order formula.
   nsubsteps = 2
!  How many t-steps between each value of t at which
!  the fields are written out:
   nt_writeout = 1

!  Choice of integration rule for the integration in the 
!  z-direction, propagation distance (in mum) and number
!  of steps:
   izrule = 3
   zmax = 16.d0
   n_z_steps = 1600
!  How many z-steps between each value of z at which
!  the fields are written out:
   nz_writeout = 20

   density = 1.96d+21  ! 220 C.

   Gamma_decay_f(1,2) = 5.746d0
   Gamma_decay_f(1,3) = 6.0666d0
   detuning_fact(2,1) = -1.d0
   detuning_fact(3,2) = -1.d0
   detuning(1) = 0.d0
   detuning(2) = 0.d0
   dip_mom(2,1,1) = 1.465d-29
   dip_mom(3,1,2) = 2.06937d-29
!  Both fields propagate in the positive z-direction:
   idir(1) = 1
   idir(2) = 1
   wavelength(1) = 794.979d0
   wavelength(2) = 780.241d0

!  The following information determines the initial
!  density matrix as explained in the User Manual.
   istart = 2
   popinit(1) = 1.d0

!  The calculated fields must be written on a file called 
!  outamplitudes.dat, which, if already existing, will be
!  overwritten by the program:
   iappend = 0
   filename_tdamps_out = 'outamplitudes.dat'
\
\end{verbatim}
\end{footnotesize}

The {\tt examples} directory also contains a bespoke program doing the same calculation, although with the fields calculated directly within the {\tt mbe} module rather than read from file.


\section{Running CoOMBE using a container image}
\label{appendix:podman}

This code can be compiled and run using a container image such as one managed by the Podman tool \cite{podman}. An advantage of this method is that the user does not need to worry about installing a Fortran compiler or the required numerical libraries. This approach is becoming increasingly conventional across modern software development. The user wishing to use Podman will need to install this program for their respective operating system (this program is freely available at \cite{podman}). Once done, the user can then easily build the relevant image of the code and run the program.

We provide a Podman implementation for each of the worked examples in the {\tt examples} folder, namely a Containerfile (here a Dockerfile), a Makefile and a {\tt .sh} shell script.  
Building
the image is done by the following command:
\begin{lstlisting}
podman build -t coombe .
\end{lstlisting}
(The {\tt ldbl.f90}, {\tt obe.f90}, {\tt mbe.f90} and {\tt driveall.f90} files need to be first copied into the working directory as necessary, together with the relevant {\tt general\_settings.f90} file, data files, Dockerfile, Makefile and shell script.)
Once the image has been built, the program can be run using the command
\begin{lstlisting}
podman run -v ./:/home/coombe coombe  
\end{lstlisting}
Changing any input parameters normally requires to rebuild the image by using the {\tt podman build} command again; however, the rebuild process is typically faster than the initial rebuild.







\end{document}